\documentclass[letterpaper]{ar-1col} 
\usepackage{url}
\usepackage[dvipsnames]{xcolor}
\usepackage{graphicx,amsmath,bm}
\usepackage[hidelinks]{hyperref}

\usepackage[a4paper, left = 4 cm, right = 3 cm, top = 3 cm, bottom = 4 cm]{geometry}
\jname{Annual Review of Chemical and Biomolecular Engineering}
\jvol{AA}
\jyear{2023}
\doi{10.1146/((please add article doi))}

\begin{document}
\sloppy
\markboth{Challita et al.}{Fluid ejections in nature}

 \title{Fluid ejections in nature}

\author{Elio J. Challita,$^{1,2}$ Pankaj Rohilla,$^1$ and M. Saad Bhamla$^3$
\affil{$^1$School of Chemical \& Biomolecular Engineering, Georgia Institute of Technology, 311 Ferst Drive NW, Atlanta, GA, 30332, USA; email: saadb@chbe.gatech.edu}
\affil{$^2$George W. Woodruff School of Mechanical Engineering, Georgia Institute of Technology, 801 Ferst Drive NW, Atlanta, GA, 30318, USA}}

\begin{abstract}
From microscopic fungi to colossal whales, fluidic ejections are a universal and intricate phenomenon in biology, serving vital functions such as animal excretion, venom spraying, prey hunting, spore dispersal, and plant guttation. This review delves into the complex fluid physics of ejections across various scales, exploring both muscle-powered active systems and passive mechanisms driven by gravity or osmosis. We introduce a framework using dimensionless numbers to delineate transitions from dripping to jetting and elucidate the governing forces. Highlighting the understudied area of complex fluid ejections, this work not only rationalizes the biophysics involved but also uncovers potential engineering applications in soft robotics, additive manufacturing, and drug delivery. By bridging biomechanics, the physics of living systems, and fluid dynamics, this review offers valuable insights into the diverse world of fluid ejections and paves the way for future bioinspired research across the spectrum of life.

\end{abstract}

\maketitle

\tableofcontents



\section{Overview and Motivation}
\begin{quote}
    \textit{``Water is the driving force of all nature."}  
- Leonardo Da Vinci \cite{pfister2009leonardo}
\end{quote}

Fluid transport is a biological imperative. Living organisms pump fluids, either internally or externally, to conduct essential functions vital to their survival and proliferation. This capacity has evolved in diverse ways across kingdoms and taxa, encompassing plants, fungi and animals. Fluid ejection, the main focus of this review, is a distinct yet integral part of fluid transport  \cite{vogel1996life,marusic2021leonardo}. Despite having important ecological, morphological, and evolutionary consequences, it has received limited attention \cite{challita2023droplet,yang2014duration,Weiss}.

The complexity and diversity of fluid ejections across organisms stem from an intricate interplay between fluid dynamics and biological function. Each species has evolved unique mechanisms to utilize fluid ejections for particular needs. These mechanisms are bounded by the physical laws governing the world they inhabit and the properties of the fluids they handle.

 \begin{figure}[h!]
 \includegraphics[width=5.00in]{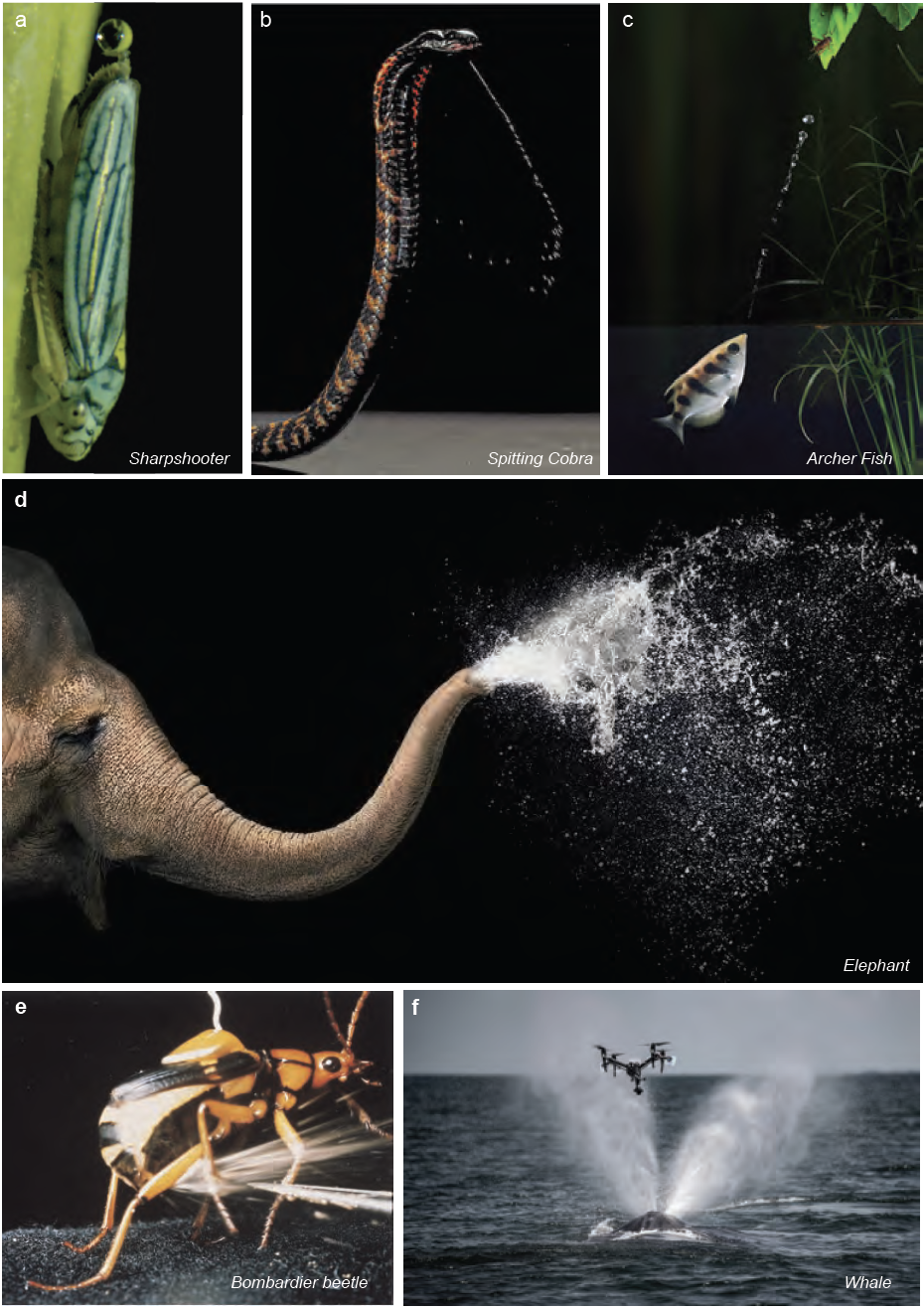}

 \caption{\textbf{Fluid ejections in nature}. Fluid ejection spans multiple scales and species, taking different forms for various functions. (\textit{a}) A sharpshooter insect excreting droplets, one at a time. (Reproduced from \cite{challita2023droplet} - Permission pending) (\textit{b}) A ring-necked spitting cobra spitting a stream of venom in self-defense (Reproduced from the Trustees of the Natural History Museum, London; Callum Mair). (\textit{c}) Archerfish expel a fluid stream to hunt insects lying on leaves. (Credit: Scott Lindstead) (\textit{d}) Elephants spraying water out of their trunks for bathing or cooling. (Reproduced with permission from Tim Flach) (\textit{e}) Bombardier beetles expelling a spray of chemicals for self-defense. (\textit{f}) Humpback whales expelling a burst of air and water jet through their blowhole as part of their breathing process. A `Snotbot' drone \cite{keller2019snotbot} collects fluidic samples of a humpback whale's liquid spout. (Panel reproduced with permission from Ocean alliance \cite{keller2019snotbot})}
 \label{fig1}
 \end{figure}


In this review, we develop a framework for understanding fluid ejections across various organisms and contexts (Figure~\ref{fig1}).  While we present a wide array of biological fluid ejections, our goal is not to provide a comprehensive overview.  Instead, we focus on rationalizing the sheer diversity of fluidic ejections in terms of 1) the organisms involved (from fungi to whales); 2) the biological functions served (excretion, hunting, defense, predation, dispersion, etc.); 3) the mechanisms utilized (catapulting, pumping, destructive); 4) their physical and temporal scales (microns to meters, milliseconds to seconds); and (5) fluid characteristics (viscoelasticity, adhesion, chemical reaction etc.). We pay particular attention  to  how biological organisms contend with  fluidic forces across physical scales, concentrating on identifying dominant forces through non-dimensional numbers and analyzing how these forces shape the ejected fluids. In some cases, we offer preliminary quantitative and order-of-magnitude analyses and hypotheses in areas where fluid ejections are understudied or not fully understood. We exclude certain topics like anatomy and the evolutionary history of fluidic ejections from the scope of this review. We hope that biologists, especially those interested in organismal biofluid transport, as well as physicists and engineers seeking to apply these principles to engineering challenges, will find this examination resonant.

 \begin{figure}[h!]
 \includegraphics[width=5in]{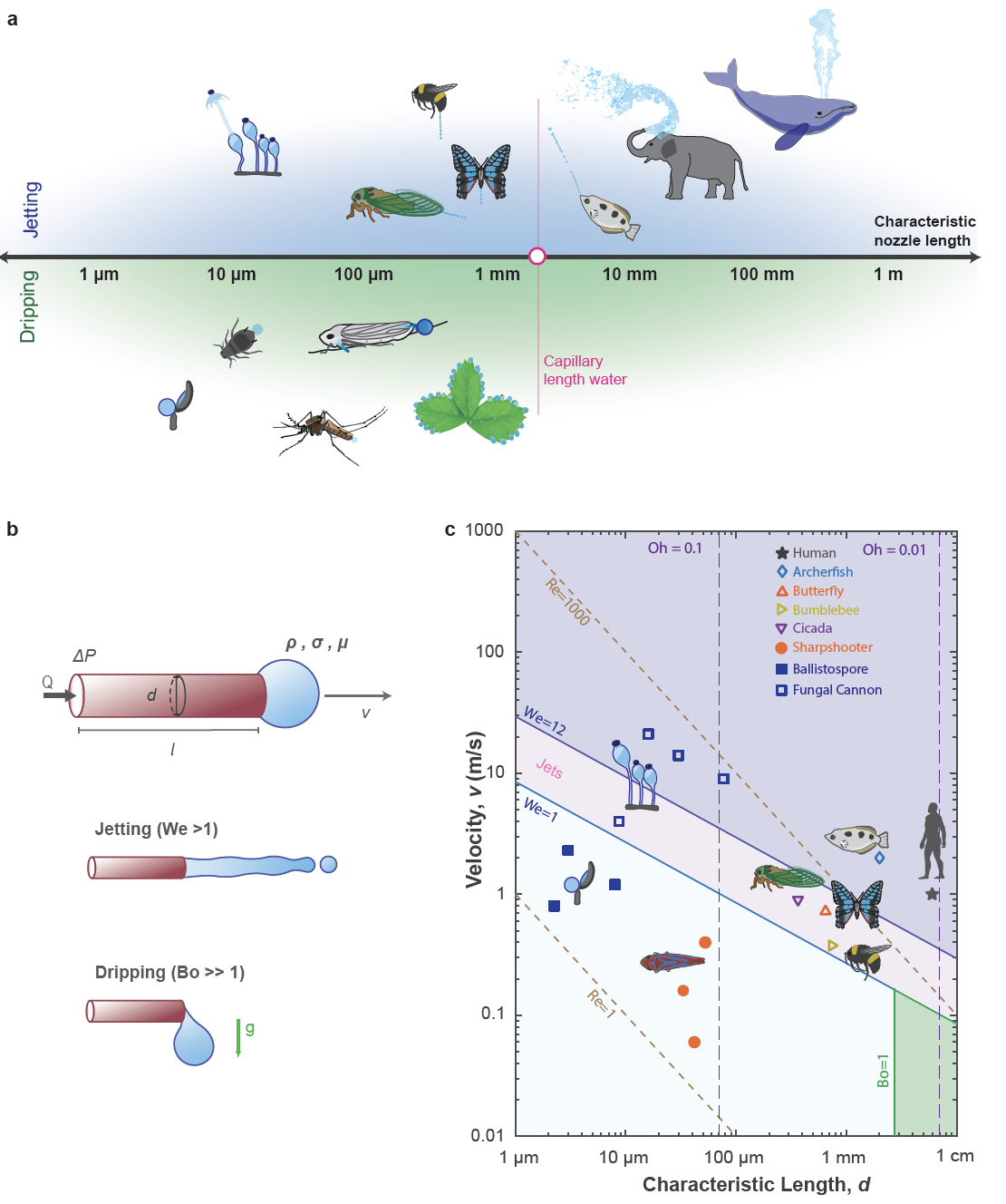} 
 \caption{\textbf{Biological Fluid Ejection: Dripping to Jetting.} (\textit{a}) Biological organisms across various taxa and length scales in nature eject fluids, which can be categorized as dripping or jetting.  (\textit{b}) The fluid dynamics of fluid ejection is evaluated by considering the fluid flow of a Newtonian fluid within a smooth, rigid, cylindrical pipe. The fluid exhibits dripping when surface tension forces are dominant $Bo<1$ $We<<1$ and exhibit Jetting when inertial forces or gravitational forces at high $We>>1$. (\textit{c}) Considering water as the fluid flow with a speed $v$ out of an orifice with a characteristic length (diameter) $d$. The lines correspond to the following dimensionless numbers: Bond number $Bo=\rho g d^2/\gamma$, Weber number $We=\rho v^2 d/\gamma$, Ohnesorge number $Oh = \mu / \rho d \gamma$, Reynolds number $Re=\rho u d/\mu$,  reflecting the interplay of surface tension, gravitational, inertial, and viscous forces. The highlighted areas reflect the various fluid regimes, including dripping and jetting. }
\label{fig:fig2}
 \end{figure}

\section{Fundamentals of Newtonian Fluid Ejection}

Fluid ejection describes the process where a  stationary fluid within a pipe begins to move and eventually exits through an orifice (nozzle). The shape of the exiting fluid depends on the dominant hydrodynamic forces, fluid properties, flow speed, and geometry/mechanical properties of the pipe/nozzle. The study of fluid dynamics of fluid ejections has been a subject of fascination throughout history, captivating scientists and engineers alike (Figure S2). This study dates back to Leonardo Da Vinci's examinations of jet behavior and fluid cohesion's role in drop formation (Figure S1)\cite{da2012notebooks}. Recent studies have further outlined fluidic ejections fundamentals through an orifice, focusing on jetting and breakup phenomena across scales \cite{eggers2008physics, Breakup,clanet_lasheras_1999}. 


\subsection{Fluidic Forces and Dimensionless Analysis}
Consider a Newtonian fluid of density $\rho$, with viscosity $\mu$, flowing through a circular cylindrical nozzle of length $l$ and a diameter $d$ at velocity $\textbf{u}$, under gravitational acceleration $g$. We assume a rigid, smooth cylinder with no corrugations and stagnant surrounding air (no wind flow or external perturbations) (Figure~\ref{fig:fig2}b).

The Navier-Stokes equation, essentially the continuum version of Newton's second law of motion, describes the fluid's motion:
\[ \rho \left( \frac{\partial \textbf{u}}{\partial t}+ u.\bm{\nabla} u\right )=-\nabla p + \mu \nabla^2\textbf{u} + \textbf{F}_{external} \]

Within the nozzle,  inertial forces arise from the fluid's mass and acceleration, and viscous forces relate to the fluid's inherent resistance to flow and friction with the walls. At the fluid-air interface, surface tension $\gamma$  arises due to the imbalance of intermolecular attractions. Since our focus is on the specific air-liquid interface, we exclude complex fluid-fluid interfaces \cite{fuller2012complex}, and fluidic jets formed in other viscous fluids, such as those in underwater propulsion by insect larvae, cephalopods, and other marine creatures, from the scope of this analysis \cite{gemmell2021cool}. Surface tension forces influence droplet formation and fluid jet behavior, especially at small scales \cite{Degennes}. The normal stress balance at a free surface must be balanced by the curvature pressure associated with surface tension: $\bm{n\cdot T\cdot n}=\gamma (\bm{\nabla\cdot n})$ where $\bm{n}$ is the unit normal to the surface and $\bm{T}=-p\bm{I}+\mu[\bm{\nabla u + (\nabla u)}^T]$ is the stress tensor and $p$ is the pressure \cite{bush2010interfacial}. External forces such as gravity, represented by $\textbf{F}_{external}$, also affect  fluid flow, particularly at larger scales. In summary, fluid forces can be categorized into bulk forces (viscosity, inertia), interfacial (surface tension), and external (gravitational), i.e.,



\begin{align*}
  {\mathbf{F}}_{f} = f(\mathbf{F}_{v}, \mathbf{F}_{\mu}, \mathbf{F}_{{\gamma}}, \mathbf{F}_{{g}})
\end{align*}



To assess these forces' relative magnitude, we consider several dimensionless numbers, including, Bond number ($Bo$), Weber number (\textit{We}), Reynolds number (\textit{Re}), Ohnesorge number (\textit{Oh}), Froude number ($Fr$) in Box 1. Within our initial assumptions' context, the relative magnitudes of hydrodynamic forces determine the fluid's shape exiting a nozzle. This represents a simplified model, as many biological organisms exhibit more complex characteristics like non-Newtonian behavior, viscoelasticity, internal corrugation, non-cylindrical geometries, and flexible or active nozzles (discussed in a later section). \\

\begin{textbox}[h]
\textbf{Box 1: Dimensionless Numbers for Newtonian Liquids in Air.} The following dimensionless numbers are essential in characterizing Newtonian fluid ejection phenomena in air, where \( v \) is the 1-D fluid speed at the nozzle exit, \( d \) is the diameter of the nozzle, \( \rho \) is the fluid density, \( l \) is the length of the tube,\( g \) is the gravitational acceleration, \( \gamma \) is the surface tension, and \( \mu \) is the fluid viscosity:
\begin{align*}
\text{Bond number (Bo)}    & : \frac{\text{gravity}}{\text{surface tension}} = \frac{\rho g d^2}{\gamma} \\
\text{Weber number (We)}   & : \frac{\text{inertia}}{\text{surface tension}} = \frac{\rho v^2 d}{\gamma} \\
\text{Reynolds number (Re)} & : \frac{\text{inertia}}{\text{viscous forces}} = \frac{\rho v d}{\mu} \\
\text{Ohnesorge number (Oh)} & : \frac{\text{viscous forces}}{\sqrt{\text{surface tension} \cdot \text{inertia}}} = \frac{\mu}{\sqrt{\rho \gamma d}} = \frac{\sqrt{We}}{Re} \\
\text{Froude number (Fr)}   & : \frac{\text{inertia}}{\text{gravity}} = \frac{v^2}{gl} 
\end{align*}
\label{box:box1}
\end{textbox}





\begin{textbox}[h!]
\label{Box_2}
\textbf{Box 2: Dripping and Jetting Regimes for Newtonian Fluids.}
A Newtonian fluid exiting a nozzle into the air can be categorized into two main regimes: dripping and jetting, characterized by the Bond number ($Bo$) and the Weber number ($We$) (Figure~\ref{fig:fig2}).

\subsection{Dripping Regime \(Bo < 1\) and \(We < 1\)}
\label{Dripping_regime}
The fluid exits in droplet form, growing slowly before detaching at a critical size. Gravity plays a significant role in droplet breakup, producing large constant-mass droplets at a slow rate (Figure~\ref{fig:fig2} a-b).

\subsection{Jetting Regime $We >1$} As the exiting fluid's speed increases, inertial forces overcome surface tension, transitioning into jetting. Various behaviors emerge, including periodic, chaotic dripping ($1<We<8$) and a continuous stream of fluid jetting at higher \(We\) values (Figure~\ref{fig:fig2} a-b).

\subsection{Breakup Regimes $We>>1$} In jetting, jet dynamics are sensitive to mechanical perturbation or thermal excitation. A small disturbance will grow when its wavelength exceeds the jet's circumference, leading to breaking up into droplets. Different jet breakup regimes can be identified on the basis of the Ohnesorge number ($Oh$) and the Weber number for liquid ($We_L$) and gas ($We_g$) \cite{DropJet}. Jet break-up regimes are not the focus of the review and are presented in Figure S3 and Supplementary Info.

\subsection{Ohnesorge number ($Oh$): Viscosity and inertio-capillary breakup}
The Ohnesorge number ($Oh$),  sheds light on the stability and breakup of a liquid jet, as well as the formation of satellite droplets. At \(Oh \ll 1\), a fluidic jet lies in the inertio-capillary regime, where it may break quickly into smaller satellite droplets due to  inertia and surface tension's combined effects. At $Oh > 1$, or at $d < 10^{-8} $m for water, viscous forces become relevant in jet dynamics.  Formed jets breakup faster due to viscous forces.

\subsection{Froude number ($Fr$): Gravity and inertia}
The Froude number ($Fr$) evaluates the relative magnitude of inertia over gravitational effects in fluid jets. A classic example of a low Froude number is the falling jet stream driven by gravity. Alternatively, a jet with high initial velocity results in a significantly high Froude number.

\end{textbox}

\begin{figure}[h!]
\includegraphics[width=0.8\textwidth]{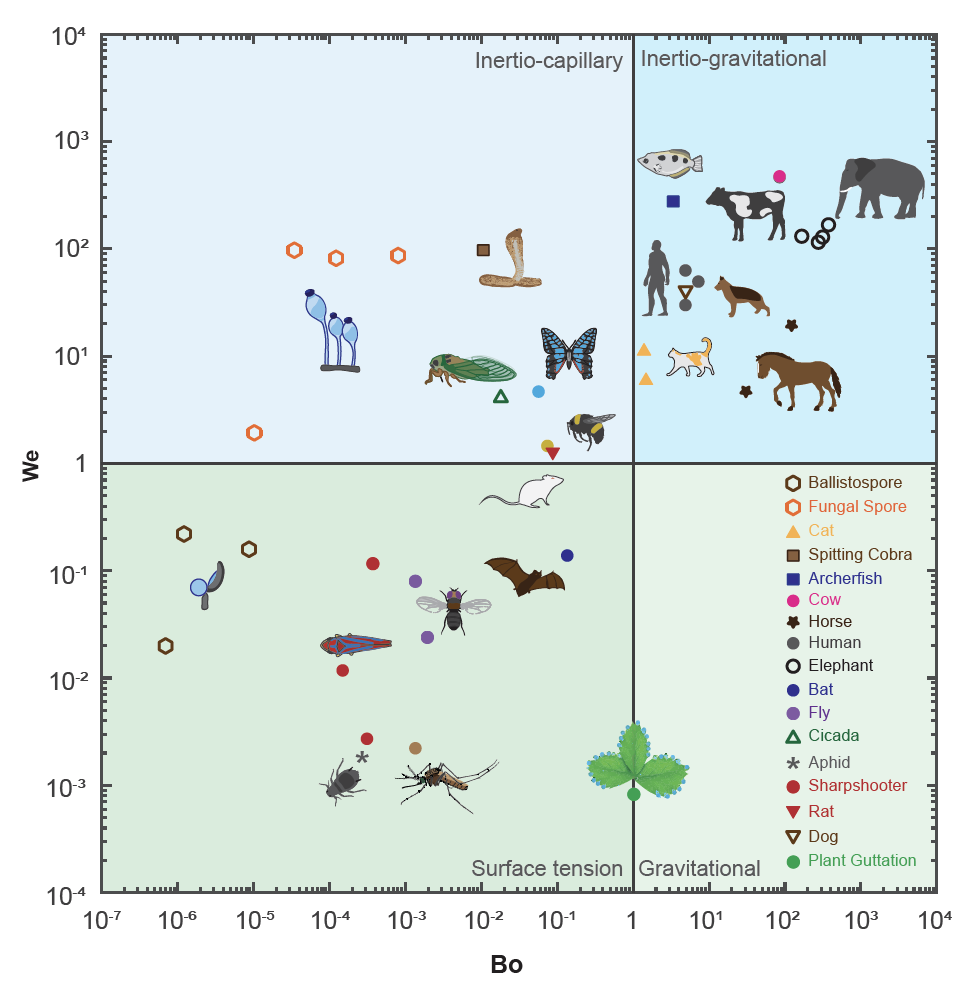}
\caption{\textbf{Mapping Newtonian Fluid Ejections via a Bond ($Bo$) and Weber ($We$) Numbers Framework.} This framework identifies four quadrants to highlight the interaction of surface tension, inertia, and gravity, ignoring the effect of viscous forces  ($Oh<<1$) - Full list in Supplementary information Table S1. In the surface tension regime ($Bo<1$, $We<1$) surface tension dominates, causing fluids to exit the nozzle as droplets. In the inertio-capillary regime ($Bo<1, We>1$), fluid ejections form jets influenced by inertial forces and surface tension. The inertio-gravitational regime ($Bo>1, We>1$) corresponds to ejections where inertia and gravity (hydrostatic pressure) cause fluids to exit as jets through a larger nozzle, as seen in most large mammalian urination. The gravitational regime ($Bo>1, We>1$)  is characterized by gravity slowly driving the dripping of a fluid. For the plant guttation, a droplet grows before reaching a critical weight before rolling over (The bond number considering the size of the droplet as characteristic length is used here)
}
\label{fig3}
\end{figure}


 

\section{Regimes of Newtonian Fluidic Ejections}

Building on the fundamental understanding of non-dimensional numbers presented in Boxes 1 and 2, we now present a comprehensive framework that leverages the Bond number (\(Bo\)) and Weber number (\(We\)) to categorize fluid ejection phenomena (Figure~\ref{fig3}). This framework provides a map comprising four distinct quadrants, each illustrating the dominance of specific forces, including inertia, gravity, and surface tension. For these calculations, we use the nozzle diameter as the characteristic length (and not the size of the organism) and the average speed of the exiting fluid. Note that since not all nozzles in biological systems are circular, the hydraulic diameter may be used to estimate the effective diameter $d=4A/P$ where $A$ is the area of the nozzle and $P$ is the perimeter of the nozzle.

The relationship between \(We\) and \(Bo\) serves as a critical tool to capture both the scale and speed of the exiting fluid. It allows us to delineate the underlying behavior, function, mechanism, and governing fluid principles that dictate how organisms eject fluids within specific \(We-Bo\) quadrants. It's essential to recognize that the demarcations at \(Bo=1\) and \(We=1\) are not rigid boundaries but rather indicative markers to underscore the prevailing forces. The actual transition between quadrants is more nuanced, with unity being presented here only as an intuitive heuristic. Some references, especially in engineering applications, define jetting when \(We>8\) \cite{DropJet}.

\subsection{Surface Tension Regime: $Bo<1$ , $We <1$  }
In this regime, surface tension reigns supreme, leading to the characteristic `dripping' of fluids. The cohesive forces of surface tension forms droplets and makes them adhere to hydrophilic surfaces through capillary adhesion \cite{Degennes,BergmanMatthewsXylem}. Such adhesion may cause droplets to remain attached to the nozzle's tip, posing dislodging challenges \cite{clanet1999transition}. This phenomenon is particularly pronounced for micro- to millimeter-scale organisms, whose size is comparable to or smaller than the capillary length \(l_c\). In this context, surface tension forces  can be substantial relative to the organism's size and weight, even to the point of trapping and endangering tiny organisms \cite{denny1993air,hu2005meniscus,ortega2022directional}.

\begin{figure}[h!]
\includegraphics[width=5in]{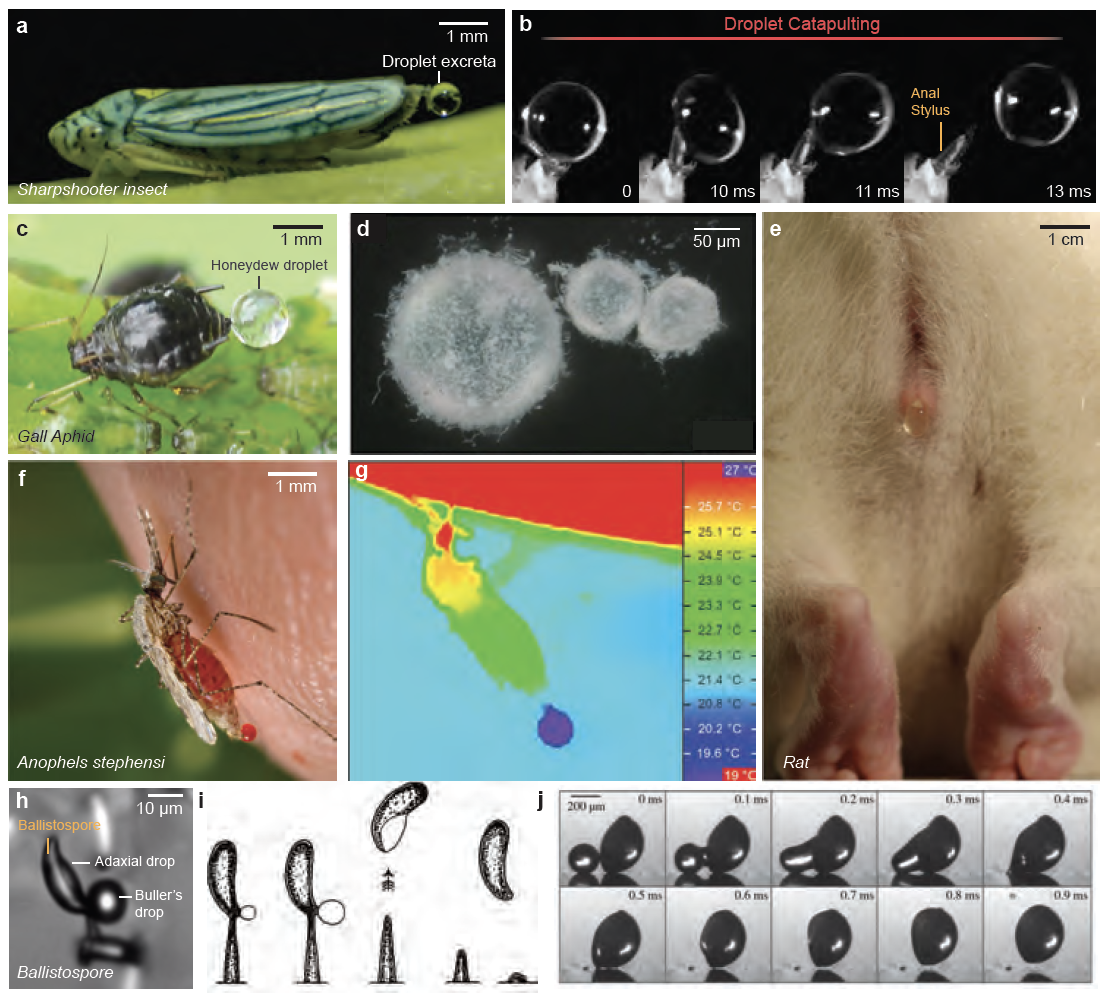}
\caption{\textbf{Exploring the Surface Tension Regime: A Diversity of Fluid Ejections.} (\textit{a}) A blue-green sharpshooter insect feeding on a basil leaf and excreting droplet excreta.  (\textit{b}) Collage showcasing the ultrafast droplet catapulting mechanism used by glassy-winged sharpshooter insects. Surface tension mediates the droplet's deformation due to its elasticity. (Reproduced with permission from \cite{challita2023droplet}) (\textit{c}) A gall-aphid excreting a honeydew droplet. (\textit{d}) Aphids secrete powdery hydrophobic wax that coats the sugar-rich droplet into a `liquid marble,' enabling easy rolling. (Reproduced from with permission from \cite{kasahara2019liquid} )  (\textit{e}) A rat urinates a droplet. (Reproduced with permission from \cite{yang2014duration})  (\textit{f}) A mosquito (\textit{Anopheles stephensi}) projecting a blood-rich droplet for thermoregulation. (\textit{g}) Thermographic image of an \textit{Anopheles stephensi} mosquito feeding on an anesthetized mouse. The mosquito cools itself through evaporative cooling, retaining a fluid drop at the abdomen's end, causing a temperature gradient along its body. (Reproduced with premissionfrom \cite{lahondere2012mosquitoes} ) (\textit{h}) Ballistospore of \textit{Tilletia caries} moments before discharge via coalescence between the Buller's and adaxial drops. (Reproduced from \cite{stolze2009adaptation}) (\textit{i}) Buller's 1922 illustration of the discharge of spore and fluid drop in \textit{Calocera cornea}. (Reproduced from \cite{buller1922researches}) (\textit{j}) Experimental demonstration of a ballistospore-inspired launch of a spore-shaped polystyrene particle from a flat superhydrophobic substrate following the coalescence of two inkjet-printed drops. (Reproduced from \cite{liu2021coalescence} )}
\label{fig4}
\end{figure}

 Various organisms have evolved innovative strategies to counter these forces. Since fluidic pumping is insufficient for droplet removal, they incorporate `secondary mechanisms'  tailored to counteract capillary adhesion. These methods exploit the inherent properties of surface tension. Some strategies, for example, leverage the elastic properties of surface tension  to use it as an engine for fluid ejection (e.g., superpropulsion in sharpshooter insects \cite{challita2023droplet} Figure~\ref{fig4} a-b, droplet coalescence in ballistospores \cite{LiuBallistospore} Figure~\ref{fig4} h-j, kicking in free aphids \cite{AphidMarble}). Others reduce the effect of surface tension altogether by coating the droplets with a hydrophobic layer (e.g., hydrophobic waxing in gall-aphids \cite{AphidMarble} Figure~\ref{fig4} c-d). Some organisms even rely on symbiotic relationships with external agents, such as ants, to remove the excreted fluids \cite{AphidMarble}. Organisms like some mosquito species exploit capillary adhesion to retain the droplets for evaporative cooling, a process that, while distinct, shares the underlying principle of evaporation with the cooling mechanism in mammals through sweating, where sweat glands release fluid that evaporates from the skin's surface to reduce body temperature \cite{folk1991evolution} (Figure~\ref{fig4} f-g).
\begin{marginnote}
\entry{Superpropulsion}{a phenomenon in which an elastic projectile (e.g., water droplet) may be ejected at a higher speed than the speed of the actuator through temporal tuning. Sharpshooter insects exploit superpropulsion of droplets to eject their excreta. \cite{challita2023dynamics} }
    
\end{marginnote}
\subsubsection{Excretion in Sap-Feeding Insects: Kicking, Catapulting, Pinching, and Waxing}
\label{sec:sec3}
Sap-feeding insects (hemipterans) are among the most extreme biological pumps \cite{BergmanMatthewsXylem}. They subsist on a liquid diet  of plant fluids, using specialized styles to extract plant sap from phloem and/or xylem tissues\cite{Redak}. With high fluid consumption rates and extended feeding periods at specific sites, they generate significant volumes of liquid waste at singular locations. Fluid ejection  becomes crucial  for  hygiene, fostering symbiotic relationships, and distancing  from waste to deter predators \cite{Weiss}.

Xylem-feeding insects like  sharpshooter insects consume vast quantities of xylem sap, approximately \(300\times\) their body weight per day \cite{Weiss}. They expel  liquid waste using a specialized appendage called the anal stylus (anal ligulae), forming a water droplet  within $80\;$ms and then catapulting it at $14\;$g \cite{challita2023droplet} (Figure~\ref{fig4} a-b).  These insects exploit the elastic properties of droplets to eject  fluidic droplets, using surface tension to elastically deform the droplets during catapulting, acting as a Hookean spring \cite{Okumura_2003, Hubert2015-sk}. By tuning the frequency of their anal stylus (the actuator) to match the Rayleigh frequency  of their elastic droplets, they efficiently remove fluidic excreta using `superpropulsion'.  \cite{challita2023droplet,Superpropulsion}. The stylus's surface also exhibits parahydrophobic properties, enabling droplets adherence for control while maintaining a high contact angle for easy removal \cite{Parahydrophobic}. 

The choice to catapult discrete water droplets instead of continuous fluid jets in sharpshooter insects is attributed to two primary energetic constraints: 1) their nutrient-scarce diet, consisting of approximately 95\% water,  and 2) the negative pressure of the xylem sap.  Consequently, sharpshooters expend significant energy to  pump this fluid using their cibarial muscles, extracting only a small amount of nutrients \cite{Andersen1989}. Such dietary and fluid mechanics considerations point to energetic constraints  related to fluidic pumping at their small scale. \cite{SmallXylem} 

The energetics of excretion can be assessed by calculating the pressure \(P\) required for pumping droplets generated by the constriction dynamics of circular muscles located in the insect's hindgut \cite{berlinhibbs1963}. To effectively eject fluid, this pressure \(P\) must overcome  combined viscous forces  (\( F_{\mu}\propto \mu l u/d^2\)), and  surface tension forces  (\(F_{\gamma}\propto \gamma/d\)). As the diameter \(d\) becomes smaller, the overall pumping pressure increases, revealing that the pressure needed for droplet formation is notably lower—specifically \(4-8\times\) less — than that  needed for a continuous jet \(We>1\).

From an energetics standpoint, jetting out fluid waste could lead to a net  energetic deficit compared to forming and catapulting droplets in these sharpshooter insects \cite{Andersen1989, challita2023droplet}. This underscores the energetic advantage of the droplet ejection mechanism over jetting, especially at microscales. However, ejecting droplets via catapulting becomes increasingly challenging as  droplet size decreases. Smaller droplets require a greater minimum acceleration for propulsion, as they are harder to deform due to higher internal pressure (\(P\propto \gamma/d\)). Therefore, the critical acceleration required for droplet detachment scales as  $D^{-3/2}$ where $D$ is the droplet diameter \cite{WaterDropsVibrating}. Such dynamics impose constraints on the capacity of biological actuators to achieve the necessary acceleration for smaller droplet detachment. This constraint may explain why sharpshooter nymphs, which are typically smaller than adults, utilize a droplet ejection mechanism based on droplet pinching, a subject currently under investigation.

Phloem-feeding insects like aphids face  different challenges. Their excreted fluid, called `honeydew', is sugar-rich (Figure~\ref{fig4} c). The accumulation of honeydew could foster detrimental fungi or parasites and  signal their location to predators\cite{fokkema1983aphid, benton1992altruistic}.  Aphids, especially those exhibiting site fidelity,  have evolved remarkable adaptations to manage their liquid waste. Gall aphids, for example, coat their honeydew in a hydrophobic powdery wax (Figure~\ref{fig4} d),  forming a `liquid marble' that rolls away from their bodies, averting build-up and ensuring hygiene \cite{AphidMarble,mahadevan2001non,benton1992altruistic}. Free-living aphids  resort to kicking fluids away \cite{broadbent1951aphid}  or  establish symbiotic relationships with ants. Spotted lanternflies,  another phloem-feeding insect species \cite{harner2022prolonged}, expel fluidic honeydew droplets using a currently-unknown catapulting mechanism. 

\subsubsection{Fungal Ballistospores: Spore Ballistics Through Droplet Coalescence}
Numerous fungal species use surface energy properties to explosively release their ballistospores (Figure~\ref{fig4} h-j). This impulsive movement is mediated by surface tension, where potential energy stored in two separate  droplets is released via  coalescence. This coalescence occurs between a spherical Buller's drop  at the spore's hilar appendix and a flattened drop on the spore's adaxial side \cite{LiuBallistospore,Pringle2005-xb}. 
Ballistospores achieve remarkable launch dynamics, with initial velocities estimated at approximately \(1\;m/s\), although their range is limited to roughly \(100\;\mu \)m, due to viscous drag. The ejection mechanisms are  single-shot and destructive, with each spore undergoing one impulsive movement for dispersion. From an engineering standpoint, coalescence-induced jumping has  potential applications in self-cleaning, anti-icing, antifrosting, and enhancing condensation heat transfer.\cite{mouterde2017merging,liu2021coalescence}.

\subsubsection{Small Mammals: Droplet Excretion Mechanisms}
\label{sec:sec311}
Fluid dynamics principles in excretion  extend to small mammals as well. The `law of urination' in mammals states that urination time is length scale-invariant, with mammals urinating in jets within $21\pm13$ seconds \cite{yang2014duration}. However, small mammals  weighing less than $<3\;$kg, such as rodents (rats and mice) and bats, typically release excrement as individual droplets or weak jets that break up immediately through the Rayleigh plateau instability ($We\sim1$). Equating the bladder pressure required  to overcome surface tension reveals a limit of $\sim$ 100 $\mu $m  orifice diameter in mammals below which excretion in continuous jets is impossible \cite{yang2014duration}. Interestingly, this constraint does not apply to arthropods such as butterflies, cicadas, and bees, which can form jets even at smaller scales (discussed below). Larger bat species, such as the flying fox, are also documented to release their urines in jets, further supporting the notion that the ability to form continuous jets is not solely determined by size but also involves other morphological and biological factors. This observation underscores deeper morphological differences between mammals and arthropods and highlights the complexity of fluid dynamics in biological systems that warrants further study.


\subsubsection{Evaporative Cooling in Arthropods: Thermoregulation Through Excreted Drops}

Thermoregulation is vital  for all organisms,  especially  for arthropods feeding on warm-blooded hosts or sugary solutions. These species have developed complex adaptations to manage excessive heat, including evaporative cooling. During this process, phase changes like evaporation require latent heat, producing a cooling effect on the evaporating fluid and its surroundings. 

Hematophagous species like mosquitoes and sandflies excrete fluid droplets during blood feeding. Anopheles mosquitoes, for example,  reduce their internal temperature by excreting and retaining fluid droplets at their abdomen's end \cite{lahondere2012mosquitoes}. The subsequent evaporation significantly cools the mosquito's body (Figure~\ref{fig4} f-g).  Species that do not retain the fluid drop appear to lack this thermoregulatory ability \cite{lahondere2012mosquitoes}.

Honeybees and bumblebees also exhibit thermoregulation through fluid ejection \cite{heinrich1979keeping,heinrich1976heat}. They regurgitate a droplet of nectar through their mouthparts, to cool their heads,  maintaining the brain's functional temperature range. This adaptation prevents overheating in hot climates or during strenuous activities like flying and foraging.

Other observed thermoregulatory mechanisms extend beyond hematophagous species. Grasshoppers feeding on succulent plants,  Sphinx moths~\cite{adams1964evaporative}, and cicadas \cite{prange1996evaporative} have been observed to employ similar strategies to exude fluids for body heat regulation, although the specific details require further investigation. In mammals, evaporative cooling is achieved through sweating, with specialized sweat glands releasing sweat through pores approximately $60-80\;\mu $m in diameter (in humans). Unlike forceful expulsion, the sweat fluid is released gradually, facilitated by the body's heat and the evaporation process itself.



\subsection{Inertio-capillary Regime: $Bo<1$ , $We >1$}

\begin{figure}[h!]
\includegraphics[width=5 in]{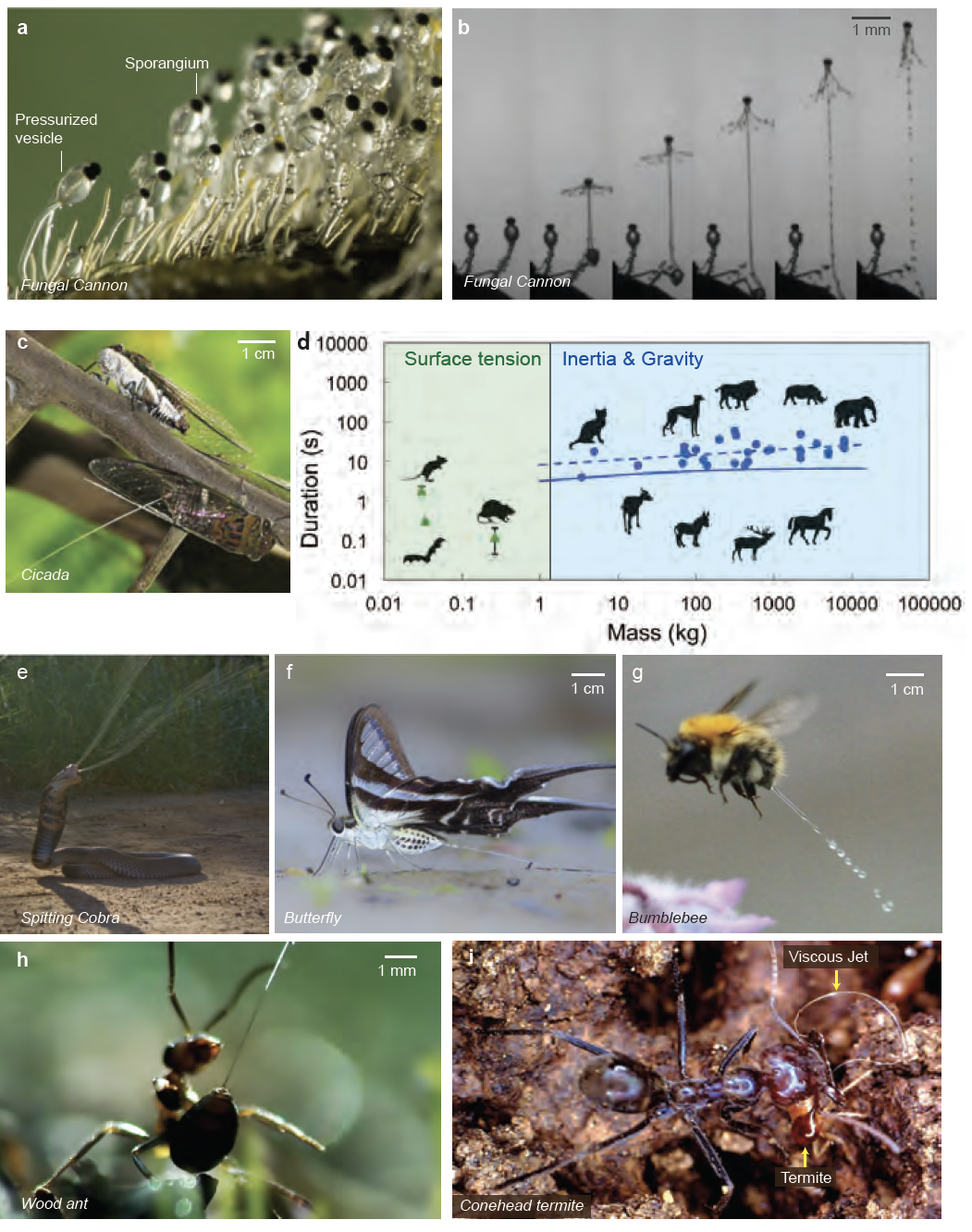}
\caption{\textbf{ Inertio-Capillary Ejections Across Species ($Bo<1$,$We<1$)}. (\textit{a}) A cluster of fungal cannons \textit{Pilobolus crystallinus} before explosively launching their spores. A pressurized vesicle generates pressure to eject the spore at up to $20,000\;$g. (Reproduced from  - Permission pending). (\textit{b}) Sequence of the spore ejection of \textit{Pilobolus kleinii} \cite{yafetto2008fastest}. (\textit{c}) Cicadas \textit{Chremistica umbrosa} feeding and peeing on Leopard tree, \textit{Caesalpinia ferrea} at the Labrador Nature Reserve, Singapore (Reproduced with permission from Tzi Ming Leong). (\textit{d}) The `law of urination' shows mammals with a mass $>3\;$kg peeing in jets within $21\pm13\;$ seconds while smaller mammals like rodents urinate in singular droplets. This law does not apply to smaller insects like cicadas, which can urinate in powerful jets despite their smaller size ($\sim 5\;$g). (Reproduced from \cite{yang2014duration}) (\textit{e}) Spitting cobra (\textit{Naja mossambica}) spitting a large venom jet in self-defense. (\textit{f}) Butterfly \textit{Lamprotera curius} emitting a stream of urine. Panel adapted from Subhendukhan (CC BY 4.0) (\textit{g}) Bumblebee emitting a fluid jet mid-air. (Credit: Mark Parrot/SWNS.com) (\textit{h}) Wood ant \textit{Formica Rufa} emitting a formic acid jet from its abdomen tip. (Credits: Terra Mater Studios) (\textit{i}) Cone-head termites \textit{Nasutitermes exositos} spitting a viscous jet against a larger ant predator (Credit: Tom Eisner).   }
\label{fig:fig5}
\end{figure}

In the inertio-capillary regime,  where $Bo<1$ and  $We >1$, inertia dominates, and surface tension plays a significant role over gravitational force (Figure~\ref{fig:fig5}). Within this regime, the Ohnesorg number (see Box 1,2) determines the dynamics of jet breakup, but we restrict our studies to the limit where $Oh<1$. For $Oh\sim 1$, the case for highly viscous fluids or very small nozzle size, fluids may become too viscous to be jettable \cite{mckinley2011wolfgang}. 

Small organisms face challenges in creating fluid jets: the pressure required to drive  fluid flow increases due to viscous forces ($F_{\text{viscous}}\propto u/d^2$) and surface tension ($F_{\text{surface tension}}\propto 1/d$) \cite{yang2014duration, challita2023droplet}. Despite these challenges, several advantages exist for small organisms in creating jets. High-speed jets enable organisms to discharge fluids at specific targets quickly, serving defensive or offensive purposes (e.g., redwood ants \cite{brutsch2017wood}, Nasutitermes \cite{challita2022viscoelastic}, and spitting cobras \cite{young2004buccal}) or  high-acceleration release of spore (e.g., fungal cannons \cite{yafetto2008fastest}). Additionally, higher-speed jets imply a higher flow rate, allowing the discharge of more fluid within a given timeframe. This capability is potentially beneficial for organisms like cicadas, which must process large volumes of low-energy nutrient xylem-sap to extract sufficient nutrients \cite{mcgavin2023essential}.

 \subsubsection{Arthropods: Jetting Chemicals for Survival}

Many arthropods, including \textit{Formica rufa} workers (redwood or horse ants), have evolved to discharge jets of chemicals for defense or predation. These ants release a precise jet of formic acid from their abdomen, covering distances of several centimeters (Figure~\ref{fig:fig5} h). This acid serves dual purposes: predation and defense, and accounts for up to 20\% of their body weight \cite{brutsch2017wood}. The production and storage of this  acid occur  in venom glands within the gaster, and is released through an `acidopore' located at the abdomen's tip. Similar behavior is observed in other ant species, such as yellow crazy ants and tawny crazy ants.  However, the fluid dynamics governing the projection of formic acid remain  not fully understood.  This lack of understanding is not surprising, considering the inherent challenges and potential dangers of filming high-speed acid jets up-close. Our personal attempts to capture these phenomena have underscored the complexity and risk involved. Given the fluid properties of formic acid ($\gamma=37\;$mN/m, $\mu = 1.78\; $mPa.s, $\rho=1220\;\text{kg/m}^3$) and the size of the acidopore  of $d<100\;\mu $m, we estimate their fluid ejection to be on the order of $We\sim \mathcal{O}(100)$.

Conehead termites (\textit{Nasutitermes}) have also evolved jet-spitting behavior for defense \cite{eisner2005secret, eisner1976defensive} (Figure~\ref{fig:fig5} j). The soldiers of these termites squirt viscous jets from a nozzle-like head projection, known as the nasus, toward perceived danger. The defensive substance serves a dual purpose: immobilizing smaller attackers and deterring  larger threats \cite{blum2012chemical}. The jets speed achieves speeds of $u\sim0.4\;$m/s within $t\sim250\;$ms, expanding almost two times their body length and forming liquid-like lassos that can physically entrap large predatory ants \cite{challita2022viscoelastic,challita2023dynamics}.

\subsubsection{Cicadas: Powerful Fluidic Jets in Xylem-sap Feeders}
Species of cicadas from the Amazon and Southeast Asia produce a range of fluidic jets. Our ongoing (unpublished) work shows that cicadas, given their larger size, have distinct fluidic capabilities compared to sharpshooters, another type of insect known for feeding on xylem-sap (see discussion in section \ref{sec:sec3}) (Figure~\ref{fig:fig5} c-d). Unlike sharpshooters, cicadas outmaneuver the inertio-capillary regime, generating fluidic jets with $We \sim 2$. This capability could be linked to their need to digest vast amounts of xylem-sap fluids to extract sufficient nutrients. However, a comprehensive understanding of the energetics encompassing feeding, nutrient extraction, and excretion in cicadas remains unexplored.  The study of cicadas has been further complicated by their varied breeding times, with some species emerging annually while others have life cycles spanning several years. The extreme difficulty in lab-rearing them has also hindered research, and to date, no one has successfully cultured cicadas in a laboratory setting \cite{challita2023cicadas}.

\subsubsection{Fungi \& Citrus: Explosive Fluid Jets}
Many fungal species, including Ascomycota, Zygomycota, and Entomophthoromycota, rely on a fracture-based release ejection mechanism to disperse their spores \cite{sakes2016shooting}.
 This process begins with water absorption into a specific chamber through osmosis, leading to an impulsive jet release when the pressure reaches a critical level ($0.31 - 1.54 \;$MPa).
 

In Ascomycota, the fluid-filled chamber,  known as asci, contains osmolytes that triggers an influx of water before discharging  with the spores.   \textit{Gibberella zeae} reports the fastest launch acceleration and velocity at $870,000\;$g and $34.5\;$ m/s, respectively \cite{yafetto2008fastest} (Figure~\ref{fig:fig5} a-b). \textit{Ascobolus immersus} and  the “spitting” fungus \textit{Glomerobolus gelineus}  achieve launch distances of approximately 0.3 m, sufficient to reach the turbulent boundary layer for wind dispersal \cite{yafetto2008fastest,kohlmeyer1996fungi}. The aerodynamic shape of the spores and the operculum minimize energy loss and drag during ejection, extending distance and reach. Ascomycetes eject spores collectively, and their synchronicity enhances dispersion, allowing spore launches to  achieve $20\times$ greater distances than individual spores \cite{van2010launched}.

\textit{Pilobolus} (Zygomycota), known as the `squirt-gun, `hat thrower', or `fungal canon', thrives on herbivore dung. It forms spore-producing structures, sporangiophores, comprising a stalk (sterigma) and a balloon-like vesicle.  The fungal jets eject the spore within $0.01$ to $0.03\;$ ms, launching it at peak accelerations of up to $21,407\;$ g and a maximum velocity of $16\;$m/s,  resulting in a launch distance of $2.5\;$m at angles ranging from $70$ to $90$ degrees to the horizontal \cite{money2009biomechanics,yafetto2008fastest}.

This fracture release mechanism is not exclusive to fungi; citrus fruits also exhibit it \cite{SmithE5887}. Oil gland reservoirs in a compressible foam-like layer near the outer surface of citrus fruits rupture under external bending deformation, raising the internal gauge pressure to  $0.03 - 0.14\;$MPa. This rupture releases high-speed microjets with a diameter of $D_0=102\pm20\;\mu $m, speeds of $8.5\pm4.0\;$m/s, and accelerations of $5100\;$g over distances exceeding $1\;$mm. These jets' characteristics are independent of the reservoir or fruit size and lead to the exhaustive ejection of the fruit's aromatic volatile oil. The citrus jets undergo different breakup regimes at $We > 8$ for speeds greater than $1.6\;$m/s then at $We>137$, corresponding to speeds over $6.6\;$m/s (see Supplementary Information S3). The function of this fluidic ejection in citrus fruits remains unknown.

\subsubsection{Cobras: Precision in Venom Spitting}  
Cobras of the \textit{Naja} genus have evolved a remarkable mechanism for defending against threats by spitting venom with high precision. These snakes release  fast, pulsed fluid through their microscopic fangs ($\sim 500\;\mu$m), directing it with high precision toward the eyes of threats up to $2\;$m away, and they can adjust the distribution of their spit with rapid movements \cite{young2004buccal,westhoff2005spitting,berthe2009spitting} (Figure~\ref{fig:fig5} e). Spitting cobras exhibit distinct morphological adaptations in their venom delivery system, setting  them apart from non-spitting species. A more circular shape of the discharge orifice, for instance, facilitates forward venom ejection, and the  venom gland's contraction provides the force needed to propel venom through the hollow fangs. Microscale venom channels, approximately $5\;$mm in length and $500\;\mu $m in diameter, feature symmetric ridges along the ventral surface, and a $90^{\circ}$ bend near the exit enhances precise control over venom flow \cite{triep20133d}. Though the venom displays shear-thinning rheology, it shows negligible non-Newtonian effects once in motion, reaching a viscosity of around $44\;$mPa.s under laminar flow conditions ($Re<100$). These physical properties ensure a predictable flow of venom. For the cobra species \textit{Naja pallida}, the ejection process consists of spits of $0.01\;$mL over a $40\;$ms duration, resulting in average speeds of approximately $1.27\;$m/s, with peak exit velocities of $5-6\;$m/s \cite{triep20133d}.

\subsection{Inertio-gravitational Regime: $Bo>1$, $We >1$}

Fluid ejection in organisms where both inertial and gravitational forces dominate the surface tension force constitutes the inertio-gravitational regime ($Bo>1$, $We>1$), where ejection is typically in the form of jets and sprays. Within this regime, the Froude number $Fr$ (see Box 1) may be relevant to measure the relative effect between inertia and gravity. 

Almost all mammals that weigh more than $3\;$ kg urinate in the form of liquid jets and sheets that are within the scope of this regime because of the high hydrostatic pressure caused by the long urethra and the larger opening size of the same. In addition to mammals, other vertebrates, such as archerfish and horned lizards, use inertio-gravitational jetting for both hunting and defending themselves from predators.  


\begin{figure}[h!]
\includegraphics[width=\textwidth]{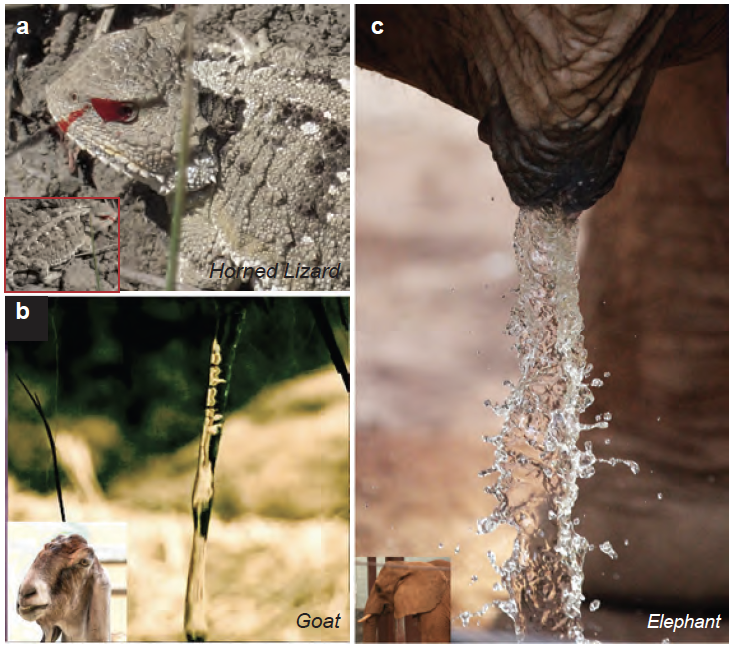}
\caption{\textbf{Jetting Phenomena in the Inertio-Gravitational Regime.} (\textit{a}) A horned lizard squirting blood from its ocular sinuses.  (Credit: J. Dullum / USFWS - CC BY 2.0 DEED) (\textit{b}) A goat's urination forms cylindrical jets (Reproduced from \cite{yang2014duration}), (\textit{c}) An elephant's urination results in chaotic ejecta sheet formation. (Reproduced from \cite{yang2014duration}).}
\label{fig6}
\end{figure}

\subsubsection{Urination in mammals} 

Mammals weighing more than 3 kg urinate in liquid jets and sheets in inertio-gravitational regime due to their large size, where inertia (and gravity) are important, while surface tension effects are negligible. The inertio-gravitational regime encompasses urination in a wide range of large mammals with $We\sim\mathcal{O}(10^{0}-10^{2})$ and $Be\sim\mathcal{O}(10^{0}-10^{2})$. The body mass of the animals in this regime ranges from $\sim5$ to $5,000$ kg, and the size of their bladders varies from 5 milliliters (ml) in cats to 18 liters (L) in elephants. Interestingly, the urine flow rate ($Q$) is proportional to the body mass ($M$) as $Q\propto M^{0.92}$ within this jetting regime (Figure \ref{fig:fig5} d).

The urination start-up phase is transient, and the jetting dynamics depend largely on the size and shape of the urethra opening and the pressure gradient. In larger animals, urination is driven by larger bladder pressure and hydrostatic pressure \cite{yang2014duration,hinman1971hydrodynamics}. Additionally, larger animals also have a longer urethra, resulting in higher hydrostatic pressure, leading to higher flow rates. Thus, larger animals expel larger amounts of liquid in a brief period of time with high flow rates, resulting in an almost constant duration across large mammals ($21\pm13$ seconds) following the `law of urination' \cite{yang2014duration} (See section~\ref{sec:sec311}, Figure~\ref{fig:fig5} d). In addition, the shape and roughness of the soft and flexible urethra opening vary in different organisms, resulting in the formation of non-axisymmetric cylindrical jets and irregular urine sheets \cite{yang2014duration,yang2023urinary,dass2001morphological} (Figure \ref{fig6} b-c). The shape factor ($4A/{\pi D^2}$) of the urethra opening is 0.14-0.22 for man to dog \cite{wheeler2012shape}, suggesting that the cross-sectional area is $\sim14-22\%$ of the cross-sectional area if it is assumed that the urethra is a rigid cylinder.

    The urination kinematics of other mammals is largely understudied, and future research should also cover marine mammals, such as blue whales, the largest mammal on Earth. Sei whales and Fin whales urinate $\sim627$ L and $\sim974$ L a day \cite{kjeld2003salt}, which is 10-20$\times$ the rate of urination in elephants. Further work is needed to address the challenges associated with the data collection of marine animals (Supplementary Information S4).


\begin{marginnote}[]
\entry{Hydrostatic pressure}{The pressure exerted by the fluid at equilibrium at any time due to the force of gravity. $P_\text{hydrostatic} = \rho g L$, where $\rho$ is the density of urine, $g$ is the acceleration due to gravity, and $L$ is the length of the urethra.}
\end{marginnote}


\subsubsection{Precision Hunting in Archerfish  Using Water Jets}


Archerfish,  belonging to the \textit{Toxotes} genus, actively hunt prey such as flies, spiders, and small lizards by shooting water jets in mangrove-filled estuaries and freshwater streams. By raising their bony tongue to the roof of their mouth, they form a tube to expel the liquid jet. The speed of the water jet at the opening of the fish mouth is $\sim2$ m/s, accelerating to $200-400$ m/s$^2$ and dropping to zero in $\sim15$ ms.  During the ballistic stage ($\sim 20-30$ ms), the jet head accelerates and increases in volume. If the tail of the jet is larger than the head, it provides additional thrust, leading to the overall axial compression of the jet. This compression determines the transfer of mass and momentum from the tail to the head of the jet (kinematic gathering phenomenon \cite{eggers2008physics}), resulting in an average impact force of $\sim200$ mN, well beyond the anchoring force of small insects and bugs. The jetting in archerfish lies in the inertio-gravitational quadrant of Figure \ref{fig3} with $We \sim \mathcal{O}(10^2$) vs. $Bo$ $\sim \mathcal{O}(10^0-10^1$).


Furthermore, a shorter impact time ($\Delta t_i \sim 1-3$ ms) of the archerfish jets allows momentum transfer. Such a short $\Delta t_i$ is enabled by the kinematic gathering and Rayleigh-Plateau instability, simultaneously amplifying the jet. For a better understanding of jet dynamics, for an orifice radius of 1 mm, the Rayleigh wavelength is $\sim9$ mm and the jet break time ($\tau_r$) is nearly 11 ms. 


As they grow and mature, the archerfish refine their jet-shooting skills for hunting, to improve the speed and the precision of their jets \cite{brodie2006watch}. They even estimate the trajectory of the prey after impact, capturing it before other predators can intervene \cite{davis2012intraspecific}. This ability is particularly impressive when considering factors such as air friction, which can vary with the prey's size and mass. The archerfish's jet amplifies with distance, a phenomenon once thought to be powered by internal structures, is actually due to the kinematic gathering and hydrodynamic instability of the jets, with the typical specific power of the jet at impact being approximately $\sim300$ W/kg -  higher than that delivered by a vertebrate muscle (500 W/kg) \cite{gerullis2014archerfish}. The archerfish regulates the liquid jet so that the front consists of a single large-volume droplet, delivering a larger momentum to the prey. By modulating the speed of the jet near the mouth, the archerfish increases the mass and speed of the jet front, generating an impact force sufficient to overcome the anchoring force of insects, typically around $\sim20$ mN (for 100 mg body mass) \cite{vailati2012archer}. 

 \begin{marginnote}[]
\entry{Kinematic gathering}{An accelerating liquid jet exiting the orifice undergoes axial compression, resulting in accumulation of liquid at the front forming a bulky head \cite{eggers2008physics} like the head formation in the jets ejected by archerfish.}
\entry{Rayleigh wavelength}{The distance at which the liquid jet becomes unstable due to Rayleigh-Plateau instability \cite{eggers2008physics}.}
\end{marginnote}

\subsubsection{Defensive Blood Squirts in Horned Lizards}

Horned lizards possess an extraordinary ability to shoot a jet of blood from their ocular sinuses (eye sockets) as a defense mechanism against predators \cite{heath1966venous, sherbrooke2001blood}. The muscles that line the veins around the eyes of the horned lizards contract, resulting in increased pressure. These lizards cause the expulsion of blood in the form of jets through further abrupt contraction of the muscles. Blood jets or squirts can travel up to a distance of around 1 m, aiming to attack predators, generally canids (dogs) \cite{middendorf1992canid}. The estimated Weber and Bond numbers for blood squirts in horned lizards are in the range of $We\sim \mathcal{O}(10^{-1}-10^{1})$ and $Bo\sim\mathcal{O}(10^{-1}-10^{1})$. These lizards also demonstrate a pattern of repetitive jetting.





\subsection{Gravitational regime: $Bo>1$ , $We <1$} 

In this regime, gravitational forces dominate over surface tension and inertial forces. This gravity-driven ejection primarily occurs either over extended periods through micrometer-scale openings (plant guttation) or slow dripping from millimeter-scale ones (dribbling).
Fluid ejection typically occurs as a result of increased forces in fluidic conduits due to high hydrostatic forces or sudden surges in internal pressure surges, such as in plant guttation. Owing to small characteristic length scales and insufficient inertia, fluid ejection may lie in a transition region from the surface tension regime to the gravitational regime. Droplets form at the conduit's tip and grow in size from the steady influx of incoming fluid. When droplets reach a critical mass where gravitational forces overcome surface tension, they detach and fall freely (Box 2). In large nozzles, a low pressure gradient in the nozzle causes droplet formation and subsequent growth, followed by free fall under the effect of gravity.

\begin{marginnote}
\entry{Plant guttation}{a process in which water droplets form at the tips or edges of leaves, often occurring when soil moisture is high and transpiration is low, usually during the night or early morning. }
\end{marginnote}

Biologically, this regime may correspond to residual fluids in a fluidic pipe or to situations where fluid control is less critical. This passive reliance on external gravitational fields can lead to challenges and potential fouling risks. Many biological fluid ejections ultimately transition to this regime, especially at the end of the ejection process, where residual fluid or muscular contraction limitations can cause a dribbling effect, e.g., dribbling after urination.

\begin{figure}[h!]
\includegraphics[width=\textwidth]{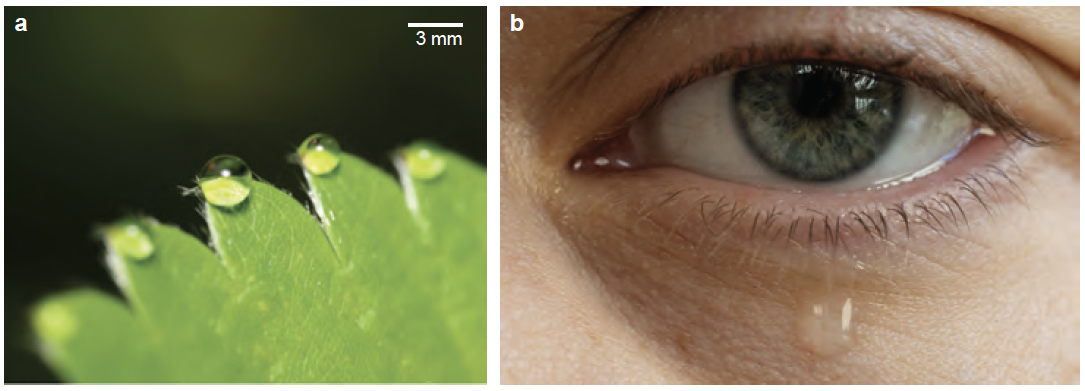}
\caption{\textbf{Gravitational Fluid Ejection in Plants and Humans.} (\textit{a}) A strawberry plant releases xylem fluid at the leaf tip after a rainy night. The droplets grow until gravitational forces overcome capillary adhesion, leading them to roll off or evaporate. (Credit: Elio J. Challita) (\textit{b}) Tears flow through openings (puncta) in the eye, and during excessive tearing, tear droplets slowly grow and drain under the influence of gravity. (Credit: Elio J. Challita) }
\label{fig7}
\end{figure}

\subsubsection{Plant Guttation} 
Root pressure releases xylem and phloem sap droplets from plant leaves through guttation.  Specialized structures known as hydathodes facilitate water exudation at the tip, edges, and margins of leaves \cite{singh2016guttation, singh2014guttation}. Droplets at the leaf apex (or edges) grow in size and wither fall under gravity or evaporate (Figure~\ref{fig7} a). Guttation mainly occurs at night when transpiration is absent. Though usually harmless, guttation can promote microbial growth, leading to plant diseases \cite{cerutti2019mangroves}. Plant guttation is one of the slowest processes of fluid ejection in comparison to almost every other example discussed in this review paper. In \textit{Zantedeschia aethiopica} plants, droplets form and fall in approximately $\sim30$ minutes, with fluid ejection charecterized by $We\sim\mathcal{O}(10^{-3})$ and $Bo\sim\mathcal{O}(10^0-10^1)$.



\subsubsection{Human Tears: Basal, Reflexive, and Emotional Tears}

Human tears, generated by the lacrimal glands, are complex physiological responses to various stimuli. Comprising three layers—an outer oily layer, a central aqueous layer, and an inner mucous layer—they nourish, lubricate, and protect the eyes. Emotional crying involves the brain's limbic system, specifically the hypothalamus, which triggers increased tear production during strong emotions. Emotional tears differ from normal ones, and tears shed by female individuals have been shown to convey a chemosignal that reduces sexual arousal and testosterone levels in men \cite{gelstein2011human}.



Tears flow through  puncta openings  ($\sim 300-400\;\mu $m in diameter) at the inner eyelid corners (Figure~\ref{fig7} b). Flow rates  range from $0.6 - 100\;\mu $L/min, depending on the tear type, whether basal, emotional, or reflex. High flow rates may cause tears  to overflow and flow down the cheek, a fluid mechanical phenomenon still understudied.

\section{Beyond the Newtonian Framework: Complex Fluid Ejections in Biological Systems}


The conventional study of fluid ejections has largely been confined to the Newtonian framework, characterized by Bond and Weber numbers (see Box 1 \& 2) (Figure \ref{fig3}), providing foundational insights into simple fluids with linear shear stress-shear rate relationships. However, biological systems often present more intricate dynamics, including non-Newtonian behaviors that transcend this elementary description \cite{ruhs2021complex}. This section marks a transition into the exploration of these complex phenomena, extending the discussion to underwater ejections, air-water interfacial dynamics, and complex airborne fluids. We delve into fluid-fluid interfacial dynamics, including Marangoni effects at the air-water interface, which were beyond the scope of the previous section's focus on aerial fluid ejection. The complexity of multiphase flows, such as atomization, aerosolization, and turbulence, is also explored, extending beyond the single-phase liquid ejection within the Newtonian framework. We also discuss the unique shapes and properties of biological nozzles, including elastic, corrugated, and soft types, which introduce elastocapillarity and elastohydrodynamic instabilities to the ejected fluid. These complexities provide a more comprehensive understanding of jets and sprays, reflecting the multifaceted nature of fluid ejections in biological systems, and set the stage for the detailed examination of complex fluids in the subsequent sections.

\begin{marginnote}[]
\entry{Non-Newtonian Fluids}{Fluids that exhibit a non-linear relationship between shear stress and strain rate, with viscosity dependent on the applied shear rate.}
\entry{Rheology}{The study of the deformation and flow of matter, encompassing both Newtonian and non-Newtonian behaviors.}
\end{marginnote}

Non-Newtonian fluids, characterized by non-linear relationships between shear stress and strain rate, offer a broader perspective that encompasses the complex behaviors found in nature. These fluids, whose viscosity depends on the applied shear rate, are employed by organisms for various functions such as locomotion, protection, hunting, and reproduction. 

Most of non-Newtonian fluids in nature are viscoelastic, possessing both viscous and elastic components. Unlike Newtonian fluids, viscoelastic fluids resist extension, leading to unique phenomena such as beads-on-a-string \cite{BOAS_Boys, BOAS_Bhat}. These complex behaviors are characterized by dimensionless numbers such as Deborah and Weissenberg numbers, defined in Box 3, and are essential in understanding viscoelastic flow \cite{pooledeborah, dealy2010weissenberg}.

\begin{marginnote}[]
\entry{Bead-on-a-string (BOAS)}{a phenomenon where the extension of certain fluids (e.g., Saliva) leads to the formation of discrete droplets (beads) separated by thin filaments (strings). }
\end{marginnote}

\begin{textbox}[h]
\textbf{Box 3: Dimensionless Numbers for Non-Newtonian (Viscoelastic) Fluids.} The following dimensionless numbers are essential in characterizing the ejection of non-Newtonian fluids:
\begin{align*}
\text{Deborah number, $De$}     = \frac{\lambda}{T}, ~~~~~~
\text{Weissenberg number, $Wi$}    = \frac{\lambda v}{L}
\end{align*}

Essentially, $De$ is the ratio of the characteristic time of the fluid ($\lambda$) and the observation time ($T$), and it determines the degree to which elasticity is expressed in response to a transient deformation. Whereas $Wi$ is defined as a product of the characteristic time of the fluid and the characteristic rate of deformation ($v/L$, where $v$ is the fluid velocity at the nozzle exit, $L$ is the characteristic length of the nozzle). 

These numbers are particularly useful for characterizing viscoelastic flows, representing the ratio of elastic to viscous forces, and correlating the fluid's relaxation time with the observation time \cite{pooledeborah, dealy2010weissenberg}. Their application in fluid ejection in living systems has been limited, but they hold the potential to gain insights of the rheological behavior of non-Newtonian fluids in nature and the industrial and natural contexts related to them.
\end{textbox}


Viscoelastic fluids are ubiquitous in nature and industry, found in organisms such as pitcher plants, deep-sea hagfish, velvet worms, and termites, as well as in industrial applications like food production, blood testing, adhesives, and paints \cite{pitcher_plants_viscoelastic, hagfish_zintzen2011, velvet_worms_jets, gilbert1985predatory, challita2023dynamics}. The transition to a more complex understanding of fluid ejections, beyond the Newtonian framework described in the section 3, holds significant relevance for real-world applications, potentially uncovering new avenues in fields such as additive manufacturing and drug delivery. The exploration of these complex phenomena underscores the need for further research and the challenges in characterizing these intricate fluid behaviors.

\subsubsection{Active Nozzle Control in Velvet Worms' Slime Jets} 
\begin{figure}[h!]
\includegraphics[width=\textwidth]{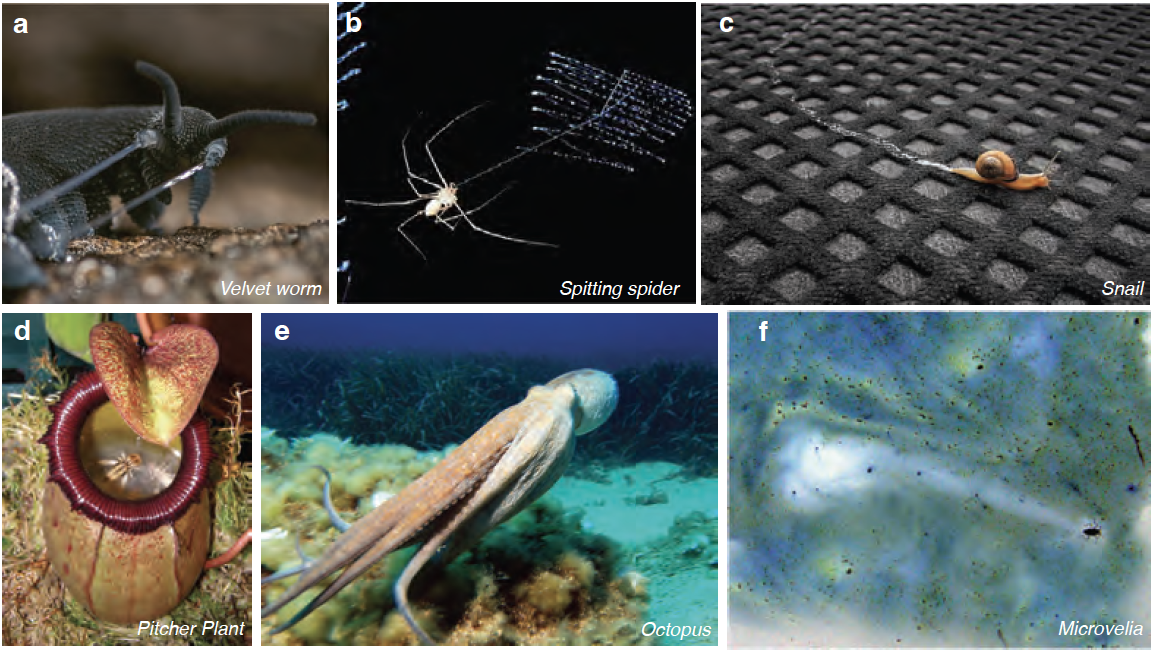}
\caption{\textbf{Examples of complex fluid ejections in nature.}  (\textit{a}) Velvet worms (\textit{Euperipatoides rowelli}) shooting viscoelastic jets for defense (Reproduced from \cite{baer2017mechanoresponsive}), (\textit{b}) Spitting spider (\textit{Scytodes thoracica}) spitting viscoelastic jets in zigzag formation (Reproduced from  \cite{suter2009spitting}) (\textit{c}) Snails (Gastropoda) leaving slime trails, which is exuded under their muscular feet for adhesive locomotion ( Credit: 	Thales/CC BY-2.0), (\textit{d}) Pitcher plant (\textit{Nepenthes sibuyanensis}) trap insects in their viscoelastic fluids , (\textit{e}) Octopus (\textit{Octopus vulgaris}) employing jet propulsion to locomote underwater in the Mediterranean Ocean, (Credit: Albert/kok - CC BY-SA 3.0), (\textit{f}) \textit{Microvelia} moving across the water surface using marangoni propulsion by ejecting surface active fluids through their proboscis, leaving a white trail by displacing the dye (Credits: John Bush and David Hu \cite{bush2006walking}).}
\label{fig8}
\end{figure}

Velvet worms (Phylum Onychophora) eject slime jets, an extraordinary example of organismal fluid ejection, by leveraging ``actively controlled soft nozzles." \cite{velvet_worms_jets, morera2010new} (Figure \ref{fig8}(\textit{a}) shows an example of slime jetting in velvet worms). Several species of these worms can be found in Southeast Asia, Central America, and South America. Slime jets display an elastohydrodynamic instability, resulting in oscillatory motion. The worms slowly contract their slime reservoir, forcing the slime through a narrow duct, and control this contraction to enhance jet speed. The ejected jets reach speeds of approximately $\sim 3-5$ m/s over a total duration of around $\sim64$ ms and oscillate at frequencies of roughly $\sim$ 30-60 Hz \cite{velvet_worms_jets}. Employed for both defensive and predatory purposes, slime forms a web-like pattern that hardens to trap and capture prey.





\subsubsection{Spitting Spiders: Oscillatory Viscoelastic Threads}
Spitting spiders (\textit{Scytodes} spp.) utilize a complex mechanism to discharge a combination of silk, venom and adhesive through their fangs, creating a distinctive zigzag pattern (Figure \ref{fig8})(\textit{b}) \cite{gilbert1985predatory, suter2005scytodes, suter2009spitting}. This pattern is formed due to the oscillatory motion of the fangs, with the spit exiting through a narrow orifice of approximately $\sim$14 $\mu$m in diameter. The ejection lasts 35 ms and reaches a maximum speed of 28 m/s, with oscillating frequency reaching up to $\sim$2 kHz, and averaging around $\sim$1 kHz.

The zigzag pattern results from the fangs' oscillation along the lateral-medial axis, coupled with a traversing motion from ventral to dorsal due to the elevation of the chelicerae. \cite{suter2009spitting}. This motion leads to the production of glue beads on silk threads during lateral-to-medial sweeps of the fangs, while medial-to-lateral sweeps produce simple fibers. This behavior, reminiscent of the beads-on-a-string phenomenon, requires further investigation, particularly as viscoelastic jetting from narrow elastic nozzles of the order of $\mathcal{O}({10^1})$ $\mu$m holds special interest for industrial applications.

The high oscillation frequencies observed in this process are likely caused by hydrodynamic forcing, generated as the fluid is ejected through the narrow elastic nozzle. Additionally, asynchronous coupling with muscles may contribute to this phenomenon, where a single muscle contraction can induce an oscillatory motion of the fangs \cite{suter2009spitting}. This intricate mechanism demonstrates the complexity of fluid dynamics in biological systems and underscores the need for detailed study and characterization.

\subsubsection{Adhesive Locomotion: Gastropods' Viscoelastic Mucus}

Gastropods (snails and slugs) utilize viscoelastic mucus for adhesive locomotion \cite{ruhs2021complex, ewoldt2007rheological}. Figure \ref{fig8}(\textit{c}) shows a slime trail left by snail. Consisting mostly of water, with glycoproteins accounting for less than 10\%, the mucus is viscoelastic and forms a thin layer about 10-20 $\mu$m thick \cite{denny1980role}. To move on this mucus film, gastropods create muscular waves along their ventral side, which are composed of alternating contractions and relaxations \cite{parker1911mechanism, lai2010mechanics}. During contraction, these waves are transmitted to the mucus layer, which undergoes large strains that disrupt the mucus network. This elastic solid-like material yields at large strain rates and behaves like a viscous fluid. During the relaxation stage, the mucus heals and reforms its gel network, allowing locomotion \cite{denny1980role}. The rhythmic discharge of mucus is controlled by the fluctuating blood pressure in the gastropod's foot through pores in the epidermis \cite{campion1961structure}.


\subsubsection{Defensive Slime Ejection: Hagfish's Rapid Response}

The hagfish (\textit{Myxinidae)}, an eel-shaped marine fish, can generate a large amount of slime in a very short period of time, releasing it from its ventral pores into the ocean \cite{hagfish_zintzen2011,martini1998ecology}. These fish dwell in the deep oceans and feed near the sea floor.  The slime ejected by hagfish comes in contact with water and undergoes physico-chemical reactions to form a fibrous gel with ultralong threads of protein and hydrated mucus, rendering it flexible for an effective defense mechanism. When provoked, hagfish eject slime, blocking the mouth and gills of their predators \cite{boni2016hagfish,ewoldt2011non} helping them escape. Rheologically, the slime exhibits elongational thickening during predator suction feeding and shear thinning during their escape.  Furthermore, hagfish slime is expected to behave as a drag-reducing agent. Some studies limited to low concentrations showed negligible drag reduction through evaluation of Wi, and future studies using higher concentrations are required in order to characterize this behavior \cite{macminn2005design}.

\subsubsection{Carnivorous Plants: Predatory Capture with Viscoelasticity}
Carnivorous pitcher plants of genus \textit{Nepenthes} secrete viscoelastic fluids to capture prey \cite{gaume2007viscoelastic}. The species depictedin Figure \ref{fig8}(\textit{d}) is \textit{Nepenthes sibuyanensis}. In \textit{N. rafflesiana}, the capture rate of flies and ants is higher (nearly 100\%) when the Deborah number ($De$) is greater than unity, i.e., $De>1$. This indicates that higher capture rates occur when the fluid's elastic forces do not have time to relax due to the insect's swimming strokes. The Deborah number is the ratio of the elastic relaxation time of the fluid to the half-period of the swimming stroke of the insects. For $De\leq0.1$, the capture rate of the flies is as low as 0\%, while the capture rate of the ants is $\sim20\%$. The lower surface tension of the fluid ($\sim60$ mN/m) also aids in sinking the prey \cite{kang2021sticky}.

\subsection{Fluid-Driven Propulsion: Navigating Underwater and Air-Water Interfaces}

Aquatic and semi-aquatic organisms often achieve rapid propulsion underwater or at the air-water interface by either ejecting fluid from their bodies or utilizing environmental fluid. This section explores various mechanisms and examples of fluid ejection for propulsion.

\subsubsection{Underwater Jet Propulsion in Aquatic Invertebrates}
Marine invertebrates such as cephalopods (octopuses, squids, and cuttlefish), jellyfish, and insect larvae, employ jet propulsion for high-speed locomotion underwater \cite{vogel1996life, vogel2014comparative, gosline1985jet, johnson1972study, gemmell2021cool}. Pulsed jetting, where the pressure created by the pulse jetting behind the organism, produces almost $30$\% more thrust than a continuous jet \cite{krueger2003significance}. Squids, chordates, octopus and jellyfish utilize this mechanism. Figure \ref{fig8}(\textit{e}) shows an example of octopus locomoting via jet propulsion under water.  Squids, for example, are among the fastest marine invertebrates, achieving maximum speeds of $5-10$ m/s through pulsatile jet propulsion \cite{vogel1987flow}. To achieve this, squids use their body cavity as a pressure chamber and a narrow tube  termed a siphon for the expulsion of water \cite{vogel1987flow,zhu2022physics}. The elasticity of the cavity wall and specific muscles power body deformation, enabling pulsatile jetting for burst propulsion \cite{zhu2022physics, ward1972locomotory}. Salps, a type of chordate, create pulse jets with speeds of approximately $3.3$ cm/s ($Re \sim 200-400$) and usually swim at a rate of 1.2 to 1.7 cm/s \cite{gemmell2021cool, sutherland2010comparative}. Jellyfish also use single jets for locomotion and predator evasion \cite{gemmell2021cool}.  

Dragonfly larvae utilize their hindguts as extreme biological pumps, ejecting repetitive jets from their anus for propulsion underwater. They modulate the orifice of the anus using an anal valve, with a reported diameter of $0.4$ mm. The velocity of the jet during propulsion typically ranges from $1-2$ m/s, with each jet lasting $100$ ms, resulting in an estimated swimming speed of 10 cm/s \cite{roh2018asymmetry, mill1975jet}.

\subsubsection{Harnessing Surface Tension: Marangoni Propulsion in Water Striders}

Water striders from genera \textit{Microvelia} and \textit{Velia} (family \textit{Vellidae}), along Rove beetles (genus: Stenus, family: \textit{Staphylinidae}), have developed a remarkable method of rapid escape from predators by utilizing Marangoni propulsion at the air-water interface \cite{linsenmair1963relaxation, bush2007integument, schildknecht1976chemical}. Figure \ref{fig8}(\textit{e}) shows \textit{Microvelia} locomoting on water using marangoni propulsion. These insects eject surface-active fluids that lower the surface tension behind them, propelling them forward. Rove beetles of the genus \textit{Stenus} employ this mechanism to achieve speeds up to $45-70$ cm/s, nearly $30\times$ faster than their normal swimming speeds \cite{linsenmair1963relaxation}. Similarly, Microvelia species use their proboscis to eject surfactants, reaching speeds up to $17$ cm/s, slightly exceeding their water-running speed \cite{nachtigall1985swimming, bush2006walking}.

\begin{marginnote}[]
    \entry{Marangoni propulsion}{A net propulsion due to an interfacial flow generated in the direction of increasing surface tension using surface-active fluids.}
\end{marginnote}


\subsection{Explosive Expulsions: Unleashing Multiphase Sprays} 
The explosive expulsion of fluids, whether through human sneezes or dolphin blowholes, showcases the intricate interplay between biomechanics and fluid dynamics. These multiphase sprays provide not only insight into biological functions but also open doors for applications in fields like medical diagnostics and marine biology. The following sections explore specific examples, shedding light on the underlying mechanisms and potential applications.
\subsubsection{Human Sneezes: Turbulent Clouds of Aerosols}
When humans sneeze or cough, they release turbulent clouds of aerosols and droplets from the nose and mouth \cite{scharfman2016visualization,bourouiba2021fluid}. This process begins with the buildup of air pressure in the lungs, which, upon sudden opening of the glottis, bursts forth as a multiphase flow. The maximum speed of this flow during a sneeze reaches around 15 m/s \cite{han2021experimental}.  These droplets often contain pathogens, leading to the transmission of respiratory diseases when spread through coughing or sneezing \cite{nunes2013dripping}.

\subsubsection{Chuffing in Dolphins and Whales: Characterizing Blowhole Ejections}

Dolphins and whales eject turbulent two-phase impulsive jets composed of air, mucus, and water, a phenomenon known as ``chuffing". Before taking a new breath, they forcefully expel these multiphase contents from their lungs \cite{fahlman2015lung, barton2020characterization}. The whale blowholes, approximately $\sim$ 5 cm in size, allow for ejections that last around $\sim$ 0.3 seconds and reach maximum speeds of up to $\sim$ 27.5 m/s \cite{APSDFD_blowholes}. Marine scientists are particularly interested in collecting mucus from these ejections, despite the challenges, as it helps gauge the stress levels of these marine mammals (Figure \ref{fig1}(\textit{f})).


\subsection{Insect Arsenal: Chemical Warfare and Fluid Dynamics}

Insects exhibit remarkable abilities in ejecting chemical sprays for defense, showcasing a complex interplay between chemistry and fluid dynamics \cite{eisner2005secret,blum2012chemical}. Bombardier beetles, European wood ants, and twostriped walkingsticks (\textit{Anisomorpha buprestoides}) employ these chemical tactics \cite{stockan2016wood, dean1990defensive}.

Bombardier beetles store a reactant solution (hydrogen peroxide, hydroquinones, and alkanes) and a solution of peroxidase and catalase enzymes in separate chambers. During ejection, these chemicals interact, facilitating an exothermic reaction that forms a mixture of quinones, ejected at a scalding 100 \(^{\circ}\)C. The beetles discharge this defensive spray as a pulsating jet at a rate of 500 pulses per second, with a typical speed of around 10 m/s and a duration of approximately 11 ms \cite{dean1990defensive,arndt2015mechanistic}.

The twostriped walkingstick, on the other hand, fends off predators by ejecting a milky white fluid up to 30-40 cm away \cite{eisner1965defensive}. Secreted from glands near the thorax and containing terpene dialdehyde \cite{meinwald1962defense}, this fluid effectively deters various organisms, including ants, beetles, and birds.

The fluid dynamics associated with these ejections are equally fascinating. For example, the bombardier beetle controls the direction of the spray discharge using the ``Coanda effect" \cite{eisner1982spray}, a phenomenon where the fluid flows along the curvature of a surface. The study of chemical warfare and fluid ejections in insects, particularly in unsteady and complex environments, remains an open and intriguing area of research, offering promising avenues for understanding biological functions and potential applications in various fields.

\subsection{Intricate Characteristics of Nature's Fluidic Conduits}
The study of fluid ejection in nature extends beyond simple geometries, delving into the complex and multifaceted world of biological conduits. The following sections explore the elasticity, corrugations, and geometry of these conduits, shedding light on the intricate dynamics that govern fluid flow in living organisms.

\subsubsection{Elasticity: Interplay of Form and Function}
Elasticity within a nozzle significantly influences jetting dynamics. Pressure-driven flow through an elastic pipe may lead to pipe expansion, altering its shape and configuration, and potentially giving rise to instabilities. This phenomenon is observed in the mammalian urethral lumen, which can enlarge during peristaltic action \cite{woodburne1972ureteral}. Velvet worms exemplify this, employing a syringe-like ejection mechanism with a flexible, accordion-shaped oral papilla to emit a slime-like jet. The resulting fluidic ejection exhibits large, rapid oscillations, enabling the coating of extensive prey surfaces. These oscillations stem from the interplay between fluid inertia and papilla elasticity, allowing high-frequency jet oscillations without the need for rapid muscle contractions or neural control \cite{concha2015oscillation}.


\subsubsection{Corrugations: Hidden Complexities}
Biological conduits often feature non-smooth interiors with surface structures, adding complexity to fluid flow. The mammalian urethra's corrugations, for example, reduce the effective pipe area and flow speed \cite{yang2023urinary,woodburne1972ureteral}. This internal roughness becomes pronounced under high Reynolds number conditions (\(Re>2300\)), where inertia plays a significant role. The interaction of the fluid with these internal structures affects the shape and breakup of the exiting jet, adding layers of complexity to the flow profile \cite{marmottant2004spray} (Supplementary Information S4).


\subsubsection{Geometry: Shaping the Flow}
The non-circular geometries of biological pipes and nozzles, including elliptical, triangular, rectangular, and tapered shapes, influence the fluid flow profile within a channel and the hydrostatic resistance. These geometrical effects may be accounted for using a correction factor \(\alpha\) \cite{mortensen2005reexamination}. Nozzle shape may also affect droplet size during dripping, with circular nozzles yielding the largest droplets at a given critical pressure. For example, droplet volume reduced by 18\% with a triangular nozzle with stretched corners compared with a circular one. Such non-circular nozzles enable more tailored fluid control by leveraging geometry which may be the case in certain biological systems \cite{chen2004optimal}.

\section{Harnessing Fluid Ejections in Living Systems: A Pathway to Innovation}

Nature's organisms have evolved sophisticated mechanisms for fluid ejection to carry out essential activities such as hunting, eating, self-defense, and excretion. Beyond the scientific curiosity to unravel these fascinating processes, there is a wealth of knowledge to be gleaned for technical applications. Drawing inspiration from natural systems has already led to innovations such as aircraft and aerial robot wing designs, gecko-inspired adhesive materials \cite{wang2020recent}, and superhydrophobic structures \cite{zhang2008superhydrophobic}. Similarly, fluid ejection systems in living organisms present a range of potential technological applications.

\begin{figure}[h!]
\includegraphics[width=\columnwidth]{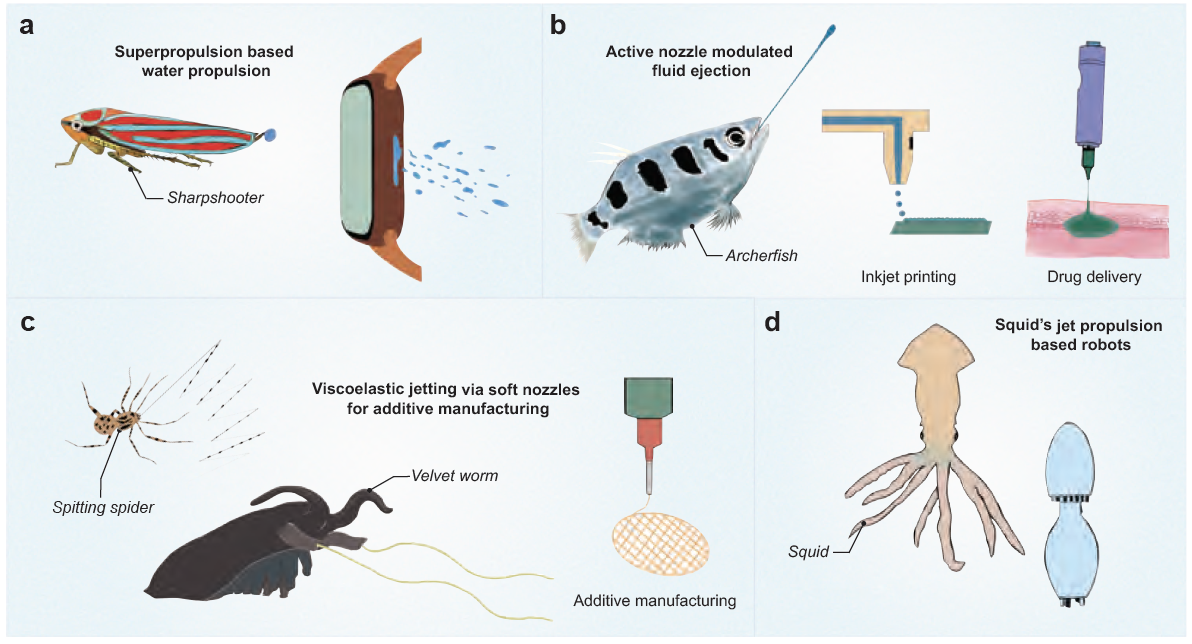}
\caption{\textbf{Innovative Applications Inspired by Natural Fluidic Ejections.} (\textit{a}) Expulsion of water from electronic gadgets like watches, utilizing the superpropulsion phenomena observed in sharpshooters  \cite{challita2023droplet}, (\textit{b})  The jetting phenomenon in archerfish, applicable in drug delivery and inkjet printing  \cite{vailati2012archer}, (\textit{c}) Micron-scale viscoelastic jetting via soft nozzles in velvet worms and spitting spiders, offering insights for nozzle design in additive manufacturing  (\textit{d})  Squid-inspired robots employing jet propulsion for rapid and efficient locomotion  \cite{xu2021squid}.} 
\label{fig:fig9}
\end{figure}



Sharpshooters launch their pee droplets using superpropulsion, an energy-efficient method of expelling liquid droplets \cite{Superpropulsion}. This mechanism could find applications in removing water from electronic gadgets, such as smartwatches, which rely on vibration-induced water removal \ref{fig:fig9}(\textit{a}). Combined with existing methods that use an acoustic signal, superpropulsion could enhance the effectiveness of water expulsion \cite{zadesky2016liquid}.

The pulsed jetting mechanisms in archerfish have potential applications in inkjet printing and drug delivery (Figure \ref{fig:fig9}(\textit{b})). Archerfish use an ``active nozzle" to increase the momentum of the jet's front, a technique that could be applied to jet injections of medications beneath the skin \cite{rohilla2019characterization}. The repetitive and continuous drop formation in archerfish jets could also be harnessed for inkjet printing.

Organisms like velvet worms, spitting spiders, and termites shoot jets of viscoelastic material through small openings (Figure \ref{fig:fig9}(\textit{c})) \cite{velvet_worms_jets, suter2009spitting}. In additive manufacturing, ejecting viscoelastic fluid through narrow micrometer-scale orifices has been challenging. Studying the nozzle structure of these organisms could guide the design of flexible and robust nozzles for applications such as 3D printing of hydrogels.




The study of ejection mechanisms and fluid dynamics in underwater organisms like jellyfish, octopuses, squids, and dragonfly larvae has inspired the development of efficient underwater robots \cite{bujard2021resonant, christianson2020cephalopod, weymouth2015ultra}. These natural propulsion systems offer valuable insights for creating innovative and effective robotic designs (Figure \ref{fig:fig9}(\textit{d})).

Other interesting applications include the development of the $\mu$Mist system by Swedish Biomimetics, which is based on flash evaporation of the liquid while exiting a chamber to form a spray and is inspired by the spray formation mechanism in bombardier beetles \cite{booth2012spray}. Furthermore, the coalescence-inducing drop jumping mechanism observed in fungal spores can be used in a wide range of applications, such as self-cleaning of surfaces, heat transfer, anti-icing, etc. \cite{mouterde2017merging,liu2021coalescence}.

\section{Concluding Remarks}
In  this review, we have dissected fluid ejections in biological systems across scales, unraveling underlying physical principles and biological functions. The intersection of biomechanics and fluid dynamics has provided insights into how organisms handle fluids, from repetitive ejections for functions like excretion and hunting to single-shot, destructive ejections in fungi and insects.

We have identified diverse mechanisms, including muscle contraction, gravity, and surface tension, leading to a wide spectrum of fluidic ejections. Large living systems rely on gravity and inertia, while small volumes depend on surface tension and inertia, with complex dynamics in both.

Beyond the Newtonian framework, we have explored intricate fluid dynamics of complex fluids, opening new avenues for technological innovation. The control over ejected fluid has been a focal point, emphasizing the balance organisms must achieve for specific functional needs. 

Open areas for research abound, from the excretion dynamics of cicadas to the chuffing of marine animals like whales. Investigating these unexplored territories may reveal insights into broader ecological patterns and environmental responses. These phenomena may serve as non-invasive `bio fluid diagnostics', offering insights into ecosystem health and the effects of climate change, akin to how human sneezing helps understand disease transmission.

The use of `active nozzles' by organisms such as conehead termites, spitting spiders, velvet worms, and archerfish to control complex fluid ejection remains an under-studied area. It holds immense potential in designing smart nozzles for applications ranging from drug delivery to on-demand printing in food engineering or additive manufacturing.

Although beyond the scope of this work, this topic is ripe for discovery. From the unexplored realms of cellular sub-cellular ejection in bacteria and protozoa to diatoms' adhesive ejection for locomotion, this review serves as a springboard to nature's immense fluid ejections waiting to be discovered. Finding the marvelous in the mundane, these phenomena open a world of exploration and inspiration.

\newpage

\begin{summary}[SUMMARY POINTS]
\begin{enumerate}
\item Nature's fluid ejections are diverse, spanning various taxa, timescales, and physical scales.\\
\item We present a framework for dripping and jetting based on the Bond (Bo) and Weber (We) numbers, categorizing fluid ejections into four quadrants: Surface tension regime ($Bo<1$, $We<1$), inertio-capillary regime ($Bo<1$, $We>1$), inertio-gravitational ($Bo>1$,$We>1$), and gravitational ($Bo>1$, $We>1$).\\
\item In the surface tension regime ($Bo<1$, $We<1$), organisms overcome fluid adhesion challenges by exploiting surface tension properties, such as ejecting droplets using elasticity, reducing it with coatings, or relying on external agents. Some organisms even leverage capillary adhesion for evaporative cooling, akin to mammalian sweating.\\
\item The inertio-capillary regime ($Bo<1$, $We>1$) sees inertial forces dominating, with surface tension more significant than gravity. Here, small organisms must overcome surface tension and viscous forces to pump fluids at high speeds. This regime includes diverse mechanisms like fracture-based ejections (fungal cannon, citrus glands) and defensive strategies (Wood ants, Nasutitermes), as well as large insects' jet urination.\\
\item In the inertio-gravitational regime ($Bo>1$, $We>1$), inertia and gravity drive fluid ejection as liquid jets and sheets. Examples include urination in large animals and the shooting of jets by archerfish and horned lizards for predation and defense. \\
\item The gravitational regime ($Bo>1$, $We<1$) involves gravity-driven fluid flow from nozzles, either as very slow dripping in small nozzles or immediate flow in large ones. Most quadrants converge into this regime at the end of their fluid ejections. \\
\item A significant subset of fluid ejection in nature falls outside the Newtonian framework of $We$ vs.\@ $Bo$. The complex dynamics of phenomena like slime ejection in terrestrial and marine organisms or marangoni propulsion in Microvelia demand further study. \\
\item The applications of fluid ejections in nature, from electronic gadgets to underwater propulsion in squids, are endless and warrant diligent exploration.
\end{enumerate}
\end{summary}

\section*{DISCLOSURE STATEMENT}
The authors are not aware of any affiliations, memberships, funding, or financial holdings that might be perceived as affecting the objectivity of this review. 

\section*{ACKNOWLEDGMENTS}
We thank Sheila Patek, Sunny Jung, David Hu, and Simon Sponberg for their valuable help and suggestions. We thank Prateek Sehgal and Jaime Quispe Nina for their help in data collection and analysis for cicada data in Peru. We thank the Bhamla lab for helpful discussions and comments.  M.~S.~B. acknowledges funding support from the NIH Grant R35GM142588 and the NSF Grants POLS-2310691, CAREER IOS-1941933, MCB-1817334, CMMI-2218382 and EF-1935262; and the Open Philanthropy Project. P.~R. acknowledges the financial support provided by the Eckert Postdoctoral Fellowship from the Georgia Institute of Technology. 

%





\bibliographystyle{unsrt}

\end{document}


\sloppy
\markboth{Challita et al.}{Fluid ejections in nature}

 \title{Supplementary Information: Fluid ejections in nature}

\author{Elio J. Challita,$^{1,2}$ Pankaj Rohilla,$^1$ and M. Saad Bhamla$^3$
\affil{$^1$School of Chemical \& Biomolecular Engineering, Georgia Institute of Technology, 311 Ferst Drive NW, Atlanta, GA, 30332, USA; email: saadb@chbe.gatech.edu}
\affil{$^2$George W. Woodruff School of Mechanical Engineering, Georgia Institute of Technology, 801 Ferst Drive NW, Atlanta, GA, 30318, USA}}



\maketitle


\begin{figure}[h!]
\includegraphics[width=6in]{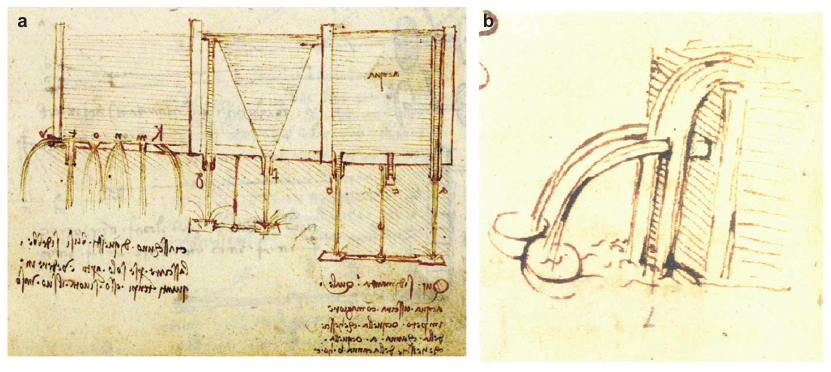}
\caption{\textbf{Leonardo Da Vinci's drawings.} (\textit{a}) ``What water will pour with the greatest violence, that from the nozzle a, b or c?'' - Leonardo Da Vinci (Reproduced from the Codex - National Library of Spain) (\textit{b}) Sketch from Leonardo's Codex on the impact of jets \cite{da1999codex,da2012notebooks,eggers2008physics} }
\label{fig:figdavinci}
\end{figure}

\begin{figure}[h!]
\includegraphics[width=\textwidth]{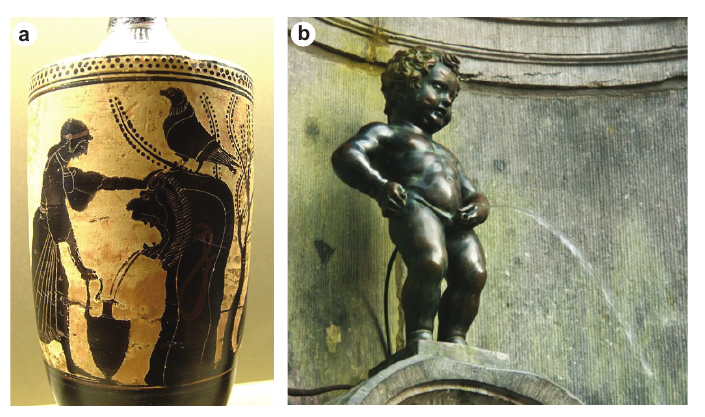}
\caption{\textbf{Historical Fountains.} (\textit{a}) Attic Greek vases from South Italy: Polyxena filling a vessel from a fountain, dated around 480 BC (Picture Source: Wikimedia, Location: The Louvre Museum, Room 652), (\textit{b}) Mannekin Pis, a bronze statue of a urinating naked young boy  (Location: Brussels, Belgium).}
\label{fig:SI_Fig_Mannekin} 
\end{figure}

\section{Historical use of fluid ejections: Water Fountains}

Throughout history, water ejection systems, such as fountains, have been used as water source for drinking, washing, and bathing in cities \cite{shakerin2000art,snoeijer2014physics, juuti2015short, shakerin2005water, hynynen2012water}. Initially, these fountains relied on gravity to create liquid jets through a narrow channel. As time passed, mechanical pumps replaced gravity as the driving force of fountains to generate powerful jets. Over time, fountains have become tourist attractions; worldwide famous example of such fountains is Manneken Pis (Brussels, Belgium) \cite{emerson2017regarding}. Fountains have been a constant witness to the advancement of science, which has enabled us to understand how to leverage the design and power to control the fluid dynamics for required applications.

\section{Active and Passive Mechanisms in Fluidic Ejection: Dynamics, Control, and Functionality}
Fluid ejection systems in nature, can be broadly categorized into active and passive mechanisms, each influencing the transient and steady-state dynamics of the fluids and setting the limits for the exerted pressure forces.

Active systems depend on muscle-powered actions, often linked to factors such as skeletal muscle force-distance relationships or the elasticity of the nozzle or pipe. Some active systems might even utilize external forces to induce a pressure differential, as seen in the milking of a cow.

In contrast, passive systems leverage external or surface forces, gravity, or osmosis to drive fluid flow, without the need for muscular action.

Control over the ejected fluid is another vital consideration. Organisms may need to manipulate the shape, speed, and breakup of the ejected fluid to meet specific functional needs such as hunting, defense, or predator avoidance. Accurate prey targeting demands high-speed fluid movement, while rapid discharge of large volumes requires maximizing the flow rate. Conversely, slow and controlled fluid release may facilitate thermoregulation through evaporative cooling. Additional complexities, such as fouling or pheromone accumulation from localized fluid concentrations, further underscore the intricate control required in fluid ejection.

\section{Jet stability}

\begin{figure}[h!]
\includegraphics[width=\columnwidth]{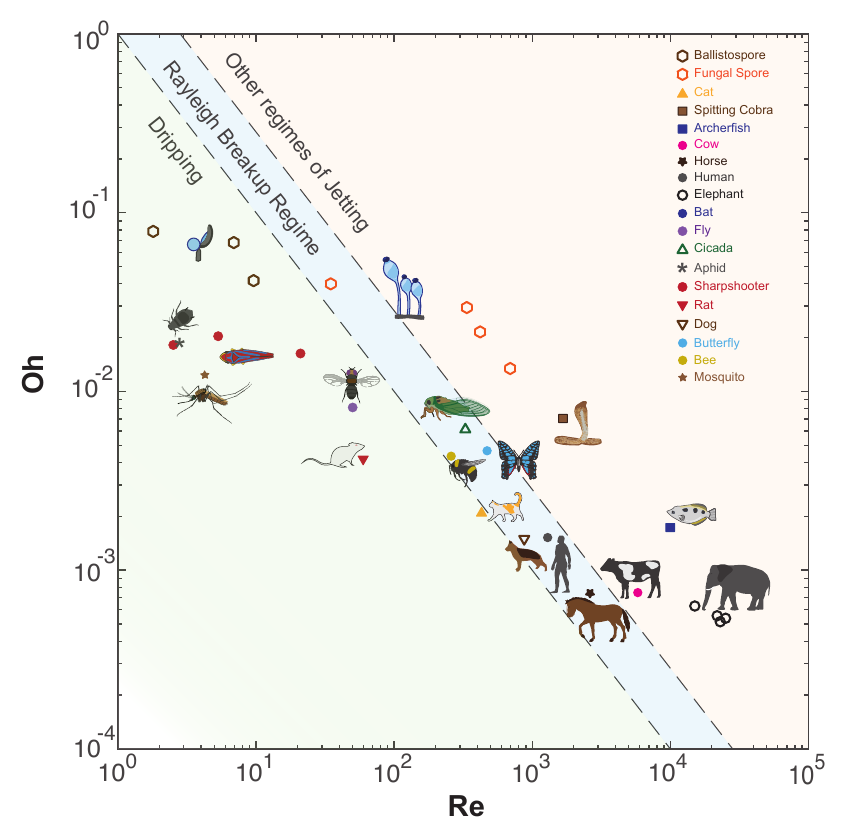}
\caption{\textbf{Ohnesorge number ($Oh$) vs. Reynolds number ($Re$) for the fluid ejections in the organisms considered in the Newtonian framework.} The dripping regime is limited to $We<1$ beyond which  jetting occurs. The jet breaks up in droplets in Rayleigh breakup regime (see Box 2), beyond which jet is continuous, yet undergoes hydrodynamics instabilities. The last regime ``Other regimes of jetting" includes other jetting regimes including first wind-induced regime, second wind-induced regime and finally atomized regime.}
\label{fig:SI_Oh_Re} 
\end{figure}

The stability of a liquid jet can be determined by different regimes within the plot of Ohnesorge number and Reynolds number. As the surrounding gas' inertial force reaches 10\% of the surface tension force, the jet becomes unstable. This is marked by the onset of the ``\textit{Rayleigh break-up regime}" ($We_L>8$ or $We_g< 1.2+3.41 Oh^{0.9}$) beyond which the ``first wind-induced" regime exists ($We_L>8$ or $We_g< 1.2+3.41 Oh^{0.9}$)  (\textit{Figure \ref{fig:SI_Oh_Re}}) \cite{DropJet}. The other two regimes of jet breakup include the second wind-induced regime ($13< We_g < 40.3$), and the atomization regime ($We_g < 40.3$). 
For simplicity, the first wind-induced regime, the second wind-induced regime, and the atomized regime are merged into a single ``\textit{Other regimes of jetting}" to show the existence of continuous jetting with instabilities. The review pertains to demarcating the dripping vs. jetting in nature, and the figure presented here (\textit{Figure \ref{fig:SI_Oh_Re}}) captures the essence of it.


As evident in this Figure \ref{fig:SI_Oh_Re} and Figure 3, ballistospores, sharpshooters, aphids, mosquitoes, flies and rats eject fluid in the form of droplets. The plot also suggests that the jets of some mammals (cats, humans, horses, cats and dogs), bees, butterfly, and cicada are unstable and break up. However, fungal pores, elephants, cobras, archerfish, and cows exhibit continuous jetting. It should be noted that the cobras and archerfish jets also break up, the deviation from the Rayleigh breakup regime could be due to variations in velocity of the jet.

\section{Other fascinating examples of Fluid ejections}

\begin{figure}[h!]
\includegraphics[width=\columnwidth]{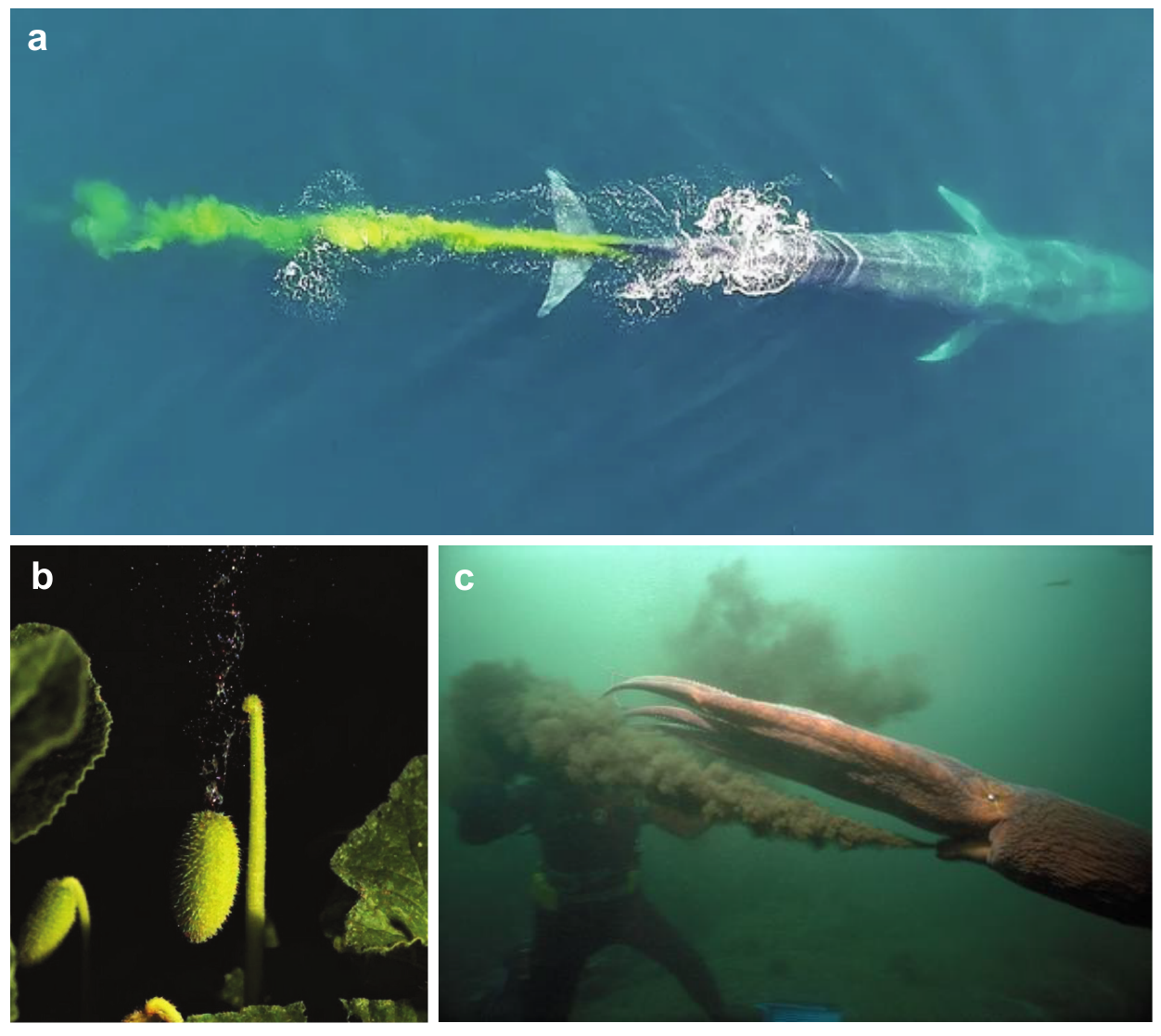}
\caption{\textbf{Some other examples of fluid ejections in organisms}.  
 (\textit{a}) Blue whale excretion in the ocean near Perth, Australia (Credit: Ian Weise, \textit{Permission pending}),  (\textit{b}) an octopus ejecting ink targeting a diver near British Columbia, Canada 
(Credit: Jeff Rotman, \textit{Permission pending}), and (\textit{c})  squirting cucumber (\textit{Ecballium elaterium}) exploding cucumber seeds (Credit: Ad{\'o}val Egy{\"u}tt) }
\label{fig:SI_Fig_whale} 
\end{figure}

Some fascinating examples of fluid ejections that are not discussed in the review are presented in Figure \ref{fig:SI_Fig_whale}. Due to limitations in data collection, some of these systems have not been studied.  

\newpage

\begin{sidewaystable}[]
\caption{Table outlining the organisms, behavior, mechanism, dimensionless numbers, and sources.}
\label{tab:tableS1}

\resizebox{\columnwidth}{!}{%
\begin{tabular}{|c|c|c|c|c|c|c|c|c|}
\hline
\textbf{Organism} &
  \textbf{Fluid} &
  \textbf{Mechanism} &
  \textbf{Behavior} &
  \textbf{\begin{tabular}[c]{@{}c@{}}Speed, \\ u (m/s)\end{tabular}} &
  \textbf{\begin{tabular}[c]{@{}c@{}}Orifice \\ diameter, d (m)\end{tabular}} &
  \textbf{$Bo$} &
  \textbf{$We$} &
  \textbf{Reference} \\ \hline
{\color[HTML]{000000} \textit{Itersonilia perplexans}} &
  Water &
  Droplet, Coalescence &
  Spore Dispersion &
  1.2 &
  8e-6 &
  8.7e-6 &
  0.16 &
  \cite{Pringle2005-xb} \\ \hline
{\color[HTML]{000000} \textit{Auricularia}} &
  Water &
  Droplet, Coalescence &
  Spore Dispersion &
  0.8 &
  2.25e-6 &
  6.87e-7 &
  0.02 &
  \cite{Pringle2005-xb} \\ \hline
{\color[HTML]{000000} \textit{Sporobolomyces}} &
  Water &
  Droplet, Coalescence &
  Spore Dispersion &
  2.3 &
  3e-6 &
  1.12e-6 &
  0.22 &
  \cite{Pringle2005-xb} \\ \hline
{\color[HTML]{000000} \textit{Ascobolus  immersus}} &
  Water &
  Jet, Fracture-release &
  Spore Dispersion &
  14 &
  3e-5 &
  1.22e-4 &
  81 &
  \cite{yafetto2008fastest} \\ \hline
{\color[HTML]{000000} \textit{Podospora anserina}} &
  Water &
  Jet, Fracture-release &
  Spore Dispersion &
  21 &
  1.6e-5 &
  3.47e-5 &
  98 &
  \cite{yafetto2008fastest} \\ \hline
{\color[HTML]{000000} \textit{Pilobolus kleinii}} &
  Water &
  Jet, Fracture-release &
  Spore Dispersion &
  9 &
  7.7e-5 &
  8.04e-4 &
  86.6 &
  \cite{yafetto2008fastest} \\ \hline
{\color[HTML]{000000} \textit{Basidiobolus ranarum}} &
  Water &
  Jet, Fracture-release &
  Spore Dispersion &
  4 &
  8.7e-6 &
  1.03e-5 &
  1.93 &
  \cite{yafetto2008fastest} \\ \hline
Glassy-winged sharpshooter &
  Water &
  Droplet, Catapulting &
  Urination &
  0.4 &
  5.2e-5 &
  3.74e-4 &
  0.12 &
  \cite{challita2023droplet} \\ \hline
Blue-green sharpshooter &
  Water &
  Droplet, Catapulting &
  Urination &
  0.16 &
  3.3e-5 &
  1.5e-4 &
  0.01 &
  \cite{challita2023droplet} \\ \hline
Red-blue sharpshooter &
  Water &
  Droplet, Catapulting &
  Urination &
  0.06 &
  4.2e-5 &
  2.4e-4 &
  0.002 &
  \cite{challita2023droplet} \\ \hline
Aphid &
  Honeydew &
  Droplet, Hydrophobic Waxing &
  Urination &
  0.07 &
  4e-5 &
  3.12e-4 &
  0.003 &
  https://tinyurl.com/279n6bb6 \\ \hline
Mosquito &
  Blood &
  Droplet, Evaportation/Squeeze &
  Thermoregulation &
  0.04 &
  1e-4 &
  0.0014 &
  0.0022 &
  https://tinyurl.com/37a6b88j \\ \hline
Fly1 &
  Assumed water &
  Droplet, Squeezing &
  - &
  0.24 &
  1e-4 &
  0.0014 &
  0.079 &
  https://tinyurl.com/mvjdxpuu \\ \hline
Fly2 &
  Asummed water &
  Droplet, Squeezing &
  - &
  0.12 &
  1.2e-4 &
  0.002 &
  0.0239 &
  https://tinyurl.com/54a697z2 \\ \hline
Cicada &
  Water &
  Jet &
  Urination &
  0.9 &
  3.64e-4 &
  0.02 &
  4.1 &
  This work \\ \hline
Butterfly &
  Assumed water &
  Jet &
  - &
  0.72 &
  6.47e-4 &
  0.06 &
  4.67 &
  https://tinyurl.com/4dw4fb8b \\ \hline
Bumblebee &
  Assumed water &
  Jet &
  - &
  0.37 &
  7.4e-4 &
  0.07 &
  1.45 &
  https://tinyurl.com/y5hdmfb4 \\ \hline
Bat &
  Urine &
  Droplet &
  Urination &
  0.1 &
  1e-3 &
  0.13 &
  0.14 &
  https://tinyurl.com/4re6vk99 \\ \hline
Wister Rat &
  Urine &
  Droplet &
  Urination &
  0.34 &
  8e-4 &
  0.09 &
  1.28 &
  \cite{yang2014duration} \\ \hline
Cat Female 1 &
  Urine &
  Jet &
  Urination &
  0.5 &
  0.0032 &
  1.39 &
  11 &
  \cite{yang2014duration} \\ \hline
Cat Female 2 &
  Urine &
  Jet &
  Urination &
  0.36 &
  0.033 &
  1.48 &
  6 &
  \cite{yang2014duration} \\ \hline
Dog &
  Urine &
  Jet &
  Urination &
  0.7 &
  0.006 &
  4.88 &
  40 &
  \cite{yang2014duration} \\ \hline
Human (Man) &
  Urine &
  Jet &
  Urination &
  0.7 &
  0.007 &
  7.23 &
  50 &
  \cite{yang2014duration} \\ \hline
Human (Woman) &
  Urine &
  Jet &
  Urination &
  0.87 &
  0.006 &
  4.88 &
  63 &
  \cite{yang2014duration} \\ \hline
Human (Old man) &
  Urine &
  Jet &
  Urination &
  0.6 &
  0.006 &
  4.88 &
  30 &
  \cite{yang2014duration} \\ \hline
Horse &
  Urine &
  Jet &
  Urination &
  0.21 &
  0.03 &
  122 &
  19 &
  \cite{yang2014duration} \\ \hline
Mini Horse &
  Urine &
  Jet &
  Urination &
  0.15 &
  0.015 &
  30 &
  4.6 &
  \cite{yang2014duration} \\ \hline
Cow 1 &
  Urine &
  Jet &
  Urination &
  1.16 &
  0.025 &
  85 &
  465 &
  \cite{yang2014duration} \\ \hline
Cow 2 &
  Urine &
  Jet &
  Urination &
  1.17 &
  0.025 &
  84 &
  473 &
  \cite{yang2014duration} \\ \hline
Elephant 1 &
  Urine &
  Jet, fluid sheet &
  Urination &
  0.43 &
  0.045 &
  274 &
  115 &
  \cite{yang2014duration} \\ \hline
Elephant 2 &
  Urine &
  Jet, fluid sheet &
  Urination &
  0.48 &
  0.05 &
  374 &
  167 &
  \cite{yang2014duration} \\ \hline
Elephant 3 &
  Urine &
  Jet, fluid sheet &
  Urination &
  0.52 &
  0.035 &
  166 &
  130 &
  \cite{yang2014duration} \\ \hline
Asian Elephant &
  Urine &
  Jet, fluid sheet &
  Urination &
  0.44 &
  0.048 &
  312 &
  128 &
  \cite{yang2014duration} \\ \hline
Spitting Cobra &
  Venom &
  Jet &
  Spitting. defense &
  5 &
  52.84e-4 &
  0.01 &
  97 &
  \cite{young2004buccal} \\ \hline
Archerfish &
  Water &
  Jet &
  Spitting, predation &
  2 &
  0.005 &
  3.4 &
  280 &
  \cite{Arch} \\ \hline
\end{tabular}%
}

\end{sidewaystable}

\bibliographystyle{unsrt}